\documentclass[sigconf]{acmart} 

\def \etal {{\emph{et al}.\thinspace}}

\usepackage{url}
\usepackage{multirow}
\usepackage{enumitem}
\usepackage{graphicx}
\usepackage{stfloats}

\newcommand{\myparagraph}[1]{\underline{#1}}

\AtBeginDocument{%
  }

\copyrightyear{2025}
\acmYear{2025}
\setcopyright{cc}
\setcctype{by-nd}
\acmConference[CHI '25]{CHI Conference on Human Factors in Computing Systems}{April 26-May 1, 2025}{Yokohama, Japan}
\acmBooktitle{CHI Conference on Human Factors in Computing Systems (CHI '25), April 26-May 1, 2025, Yokohama, Japan}
\acmDOI{10.1145/3706598.3714300}
\acmISBN{979-8-4007-1394-1/25/04}

\acmSubmissionID{7524}

\begin{document}

\captionsetup[figure]{labelfont={bf},name={Fig.},labelsep=period}
\everypar{\looseness=-1}

\title{RouteFlow: Trajectory-Aware Animated Transitions}

\author{Duan Li}
\authornote{Both authors contributed equally to this research.}
\affiliation{%
  \department{School of Software}
  \institution{Tsinghua University}
  \city{Beijing}
  \country{China}
}
\orcid{0009-0006-6941-9098}
\email{liduan429@gmail.com}

\author{Xinyuan Guo}
\authornotemark[1]
\affiliation{%
  \department{School of Software}
  \institution{Tsinghua University}
  \city{Beijing}
  \country{China}
}
\orcid{0009-0006-6399-0613}
\email{yczddgj@126.com}

\author{Xinhuan Shu}
\affiliation{%
  \department{School of Computing}
  \institution{Newcastle University}
  \city{Newcastle Upon Tyne}
  \country{United Kingdom}}
\orcid{0000-0002-9736-4454}
\email{xinhuan.shu@gmail.com}

\author{Lanxi Xiao}
\affiliation{%
  \department{Academy of Arts and Design}
  \institution{Tsinghua University}
  \city{Beijing}
  \country{China}
}
\orcid{0009-0001-5385-1453}
\email{tarolancy@gmail.com}

\author{Lingyun Yu}
\affiliation{%
  \department{School of Advanced Technology}
  \institution{Xi'an Jiaotong-Liverpool University}
  \city{Suzhou}
  \state{Jiangsu}
  \country{China}
}
\orcid{0000-0002-3152-2587}
\email{Lingyun.Yu@xjtlu.edu.cn}

\author{Shixia Liu}
\authornote{Shixia Liu is the corresponding author.}
\affiliation{%
  \department{School of Software}
  \institution{Tsinghua University}
  \city{Beijing}
  \country{China}
}
\orcid{0000-0003-4499-6420}
\email{shixia@tsinghua.edu.cn}


\begin{abstract}
Animating objects’ movements is widely used to facilitate tracking changes and observing both the global trend and local hotspots where objects converge or diverge.
Existing methods, however, often obscure critical local hotspots by only considering the start and end positions of objects' trajectories.
To address this gap, we propose RouteFlow, a trajectory-aware animated transition method that effectively balances the global trend and local hotspots while minimizing occlusion.
RouteFlow is inspired by a real-world bus route analogy: objects are regarded as passengers traveling together, with local hotspots representing bus stops where these passengers get on and off.
Based on this analogy, animation paths are generated like bus routes, with the object layout generated similarly to seat allocation according to their destinations.
Compared with state-of-the-art methods, RouteFlow better facilitates identifying the global trend and locating local hotspots while performing comparably in tracking objects' movements. 

\end{abstract}



\maketitle


\section{Introduction}

Animating objects’ movements is widely used to facilitate tracking changes and observing both the global trend and local hotspots where objects converge or diverge~\cite{cui2011textflow,tao2017hotspot}.
For example, by animating bird migration data~\cite{birdmapdemo}, users can observe birds' movements, understand the migration trend, and identify highly active locations where birds converge to cross the straits or diverge to bypass mountains (Fig.~\ref{fig:motivation}(a)).
Here, the global trend provides valuable insights into broader movement patterns, while the local hotspots serve as strategic locations for observation and analysis~\cite{li2023vectortrajectory,liu2014survey}.


\setcounter{figure}{0}
\begin{figure*}[t]
  \centering
  \setlength{\abovecaptionskip}{2.8mm}
 \includegraphics[width=\linewidth]{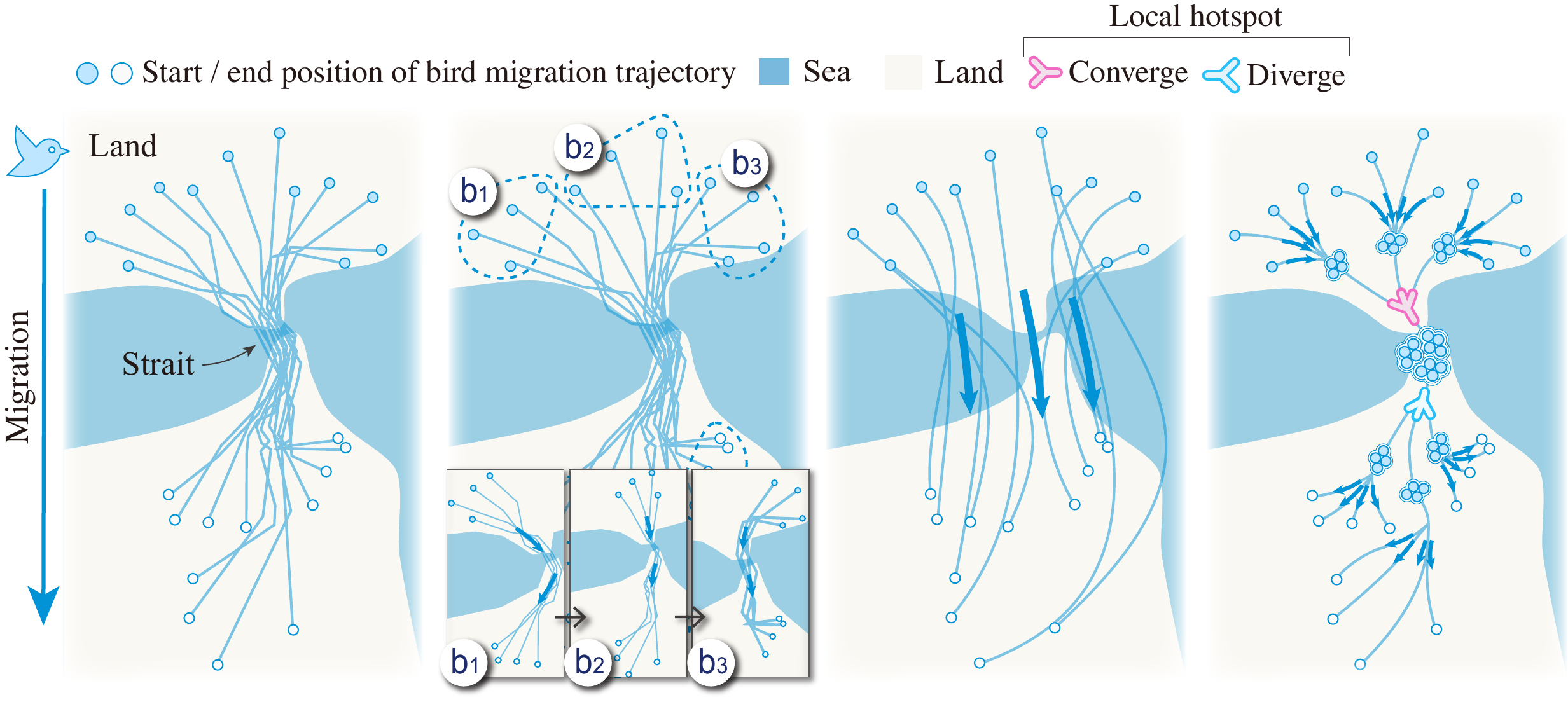}
  \put(-462,3){(a) Trajectory data}
  \put(-357,3){(b) Focus+context grouping}
  \put(-235,3){(c) Vector-field-based method}
  \put(-85,3){(d) RouteFlow}
  \caption{A comparison of three animated transition methods on a bird migration example.
  }
  \Description{A comparision of our methods with Focus+Context Grouping~\cite{zheng2018focus+}, Vector Field~\cite{wang2017vector} on a bird migration example.}
  \label{fig:motivation}
\vspace{2mm}
\end{figure*}
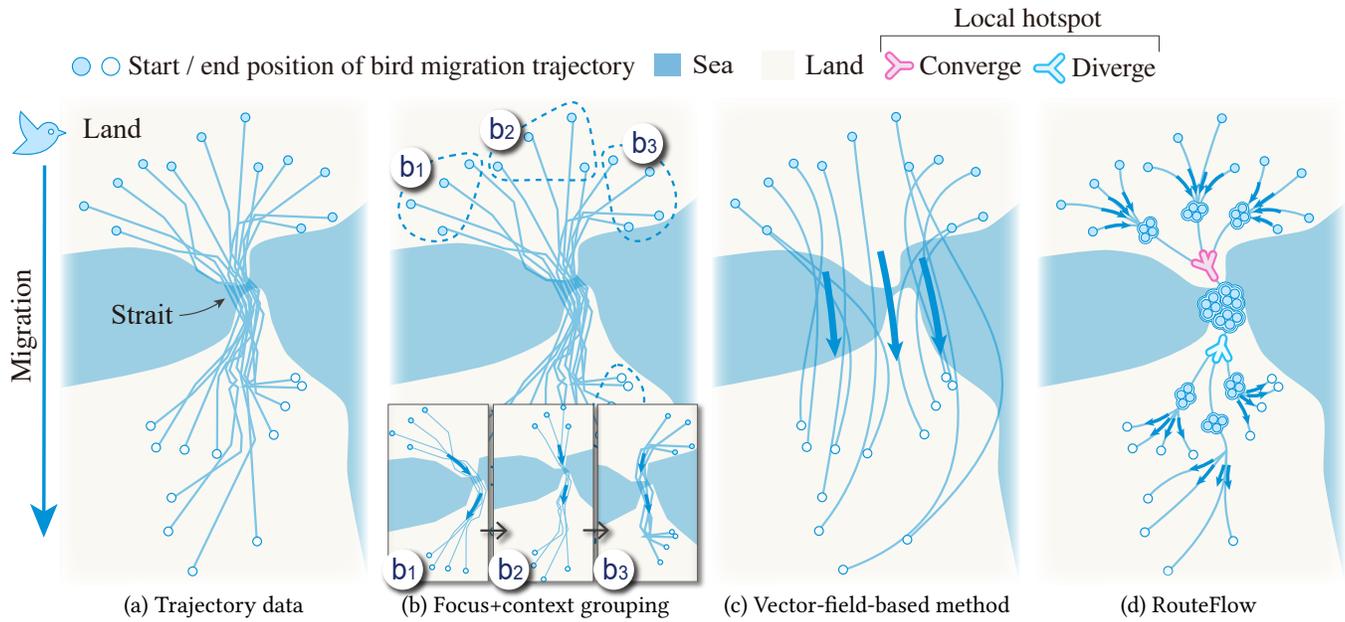

Many research efforts have been directed toward developing techniques for animated transitions, aimed at helping users track objects' movements.
These efforts mainly focus on adjusting various animation parameters from temporal (e.g., speed~\cite{dragicevic2011distortion}, staging~\cite{heer2007animated}, staggering~\cite{chevalier2014not}) and spatial (e.g., animation paths~\cite{du2015trajectory,wang2017vector}) perspectives.
Recent studies have further advanced these techniques.
Zheng~\etal~\cite{zheng2018focus+} divided transitions into groups and animated them sequentially, thereby breaking down complex animations into simpler ones (Fig.~\ref{fig:motivation}(b)).
Wang~\etal~\cite{wang2017vector} used vector fields to coordinate group movements and reduce occlusion by spatially separating animation paths (Fig.~\ref{fig:motivation}(c)).
However, all these methods only consider the start and end positions in the objects' trajectories.
Although effective in conveying global trends, they often obscure critical local hotspots along the movement trajectories.

Recognizing this gap, we aim to design an animated transition method that considers the movement trajectories of objects.
By using these trajectories, the animations can effectively reveal both the global trend and local hotspots.
Thus, our method provides a clearer understanding of local areas of high activity in their global context.
However, designing such animations is non-trivial. 
First, balancing the global trend and local hotspots in animation remains challenging. 
Overemphasizing local hotspots may result in excessive branching areas, impeding the identification of the global trend.
Conversely, stressing the global trend heavily may obscure important local hotspots.
Second, reducing occlusion in animated transitions is imperative yet difficult, especially when multiple objects move simultaneously. 
They may occlude each other, thus significantly increasing the difficulty of tracking their movements.
Occlusion becomes even more severe in local hotspots, where many objects converge or diverge. 

To address these challenges, our animation design utilizes a real-world bus route analogy: groups of passengers board the same bus at different stops, travel together along the shared routes, and disembark at designated stops.
We regard objects as passengers traveling together, with local hotspots representing various bus stops.
Based on this, we animate objects following the shared paths, converging or diverging at local hotspots. 
As such, users can observe the global trend and identify local hotspots, similar to observing overall bus routes and identifying frequently visited stops.
In this analogy, we regard 1) achieving a balance between the global trend and local hotspots as planning bus routes for efficiency and effectiveness. 
These bus routes should not only be of minimal length but also meet passengers' travel demands.
Besides, we consider 2) reducing occlusion by allocating passengers to respective seats in the process.
Consequently, we formulate the problem of designing animated transitions for trajectory data as a sequential optimization of two sub-problems: bus routing and seat allocation.

Based on this formulation, we propose RouteFlow, a trajectory-aware animated transition method comprising two steps: trajectory-driven path generation and object layout generation. 
As shown in Fig.~\ref{fig:motivation}(d), we create ``bundled'' animation paths for groups of objects that share similar movement trajectories. 
These animation paths are generated by a bottom-up hierarchical edge bundling algorithm, which progressively bundles similar trajectories, level by level, effectively capturing both the global trend and local hotspots. 
To minimize occlusion, we apply an incremental circle packing algorithm, sequentially generating the layout at each local hotspot.
The animation is then rendered using an interpolation-based method.

We evaluate RouteFlow through a quantitative experiment on real-world data and a controlled user study. 
The results indicate that compared with the state-of-the-art methods, RouteFlow better facilitates identifying the global trend and locating local hotspots while performing comparably in tracking objects' movements.
The main contributions of our work include:
\begin{itemize}
    \item A formulation of designing animated transitions as a sequential optimization of bus routing and seat allocation problems.
    \item RouteFlow, a trajectory-aware animated transition method that consists of a bottom-up hierarchical edge bundling algorithm and an incremental circle packing algorithm.
    The open-source implementation is available at \url{https://github.com/Trajectory-Anim/Trajectory-Aware-Animated-Transitions}.
    \item A quantitative experiment and a user study evaluating performance on tracking objects' movements, identifying the global trend, and locating local hotspots.
\end{itemize}



\section{Related Work}
There are two main tasks for animated transitions: \textbf{tracking objects' movements} and \textbf{identifying the trend}.
Most existing efforts focus on \textbf{tracking objects' movements} from temporal and spatial perspectives.
The temporal perspective includes adjustments such as refining movement speed~\cite{dragicevic2011distortion}, staging~\cite{guilmaine2012hierarchy,heer2007animated,zheng2018focus+}, and staggering~\cite{chevalier2014not}.
The spatial perspective focuses on the animation paths of objects~\cite{du2015trajectory,heer2007animated,wang2017vector,yee2001animated}.
Our work falls into the latter perspective.

Animation paths play a crucial role in \textbf{tracking objects' movements}~\cite{fencsik2007role,mutsumi2006trajectory}.
According to Heer and Robertson~\cite{heer2007animated}, simple trajectories are effective in minimizing confusion and enhancing predictability, thereby making it easier for users to track objects' movements.
A direct method to achieve simplicity is to use straight lines connecting the start and end positions of the movement~\cite{chang1993animation}.
To provide more natural and engaging movements while maintaining simplicity, later research used smooth curves, such as arcs~\cite{dragicevic2011gliimpse,yee2001animated} and B-splines~\cite{cao2011dicon}. 
Building on these advancements, Du~\etal~\cite{du2015trajectory} explored bundling animation paths to coordinate group movements, which improved group tracking
but could introduce the occlusion issue.
To address this, Wang~\etal~\cite{wang2017vector} utilized vector fields to coordinate movements for each group and separated the animation paths mutually to reduce occlusion among them (Fig.~\ref{fig:motivation}(c)).
However, this separation can cause much deviation from the input trajectories.
In contrast, RouteFlow creates bundled animation paths for objects with similar trajectories and reduces occlusion by applying non-overlapping constraints on the object layout.

In addition to tracking objects, animated transitions are widely used to \textbf{identify the global trend}~\cite{roberson2008trend}.
Empirical studies have discussed the potential of careful animation designs for trend identification~\cite{brehmer2020smart,chalbi2020common}. 
Recently, Zheng~\etal~\cite{zheng2018focus+} proposed the focus+context grouping method for animated transition to simultaneously track objects and identify the global trend. 
This method grouped objects with similar trends together and animated these groups sequentially (Fig.~\ref{fig:motivation}(b)).
It simplifies complex animated transitions by dividing them into a sequence of simpler groups, facilitating easier tracking of objects while also revealing the global trend.
However, this method groups transitions solely based on the start and end positions of the objects, ignoring their trajectories.
As a result, it may fail to capture important patterns throughout trajectories, e.g., the local hotspots where objects converge or diverge.
To overcome this limitation, RouteFlow considers the movement trajectories of objects, aiming to balance both the global trend and local hotspots. 



\setcounter{figure}{1}
\begin{figure*}[hbpt]
  \centering
  \includegraphics[width=\linewidth]{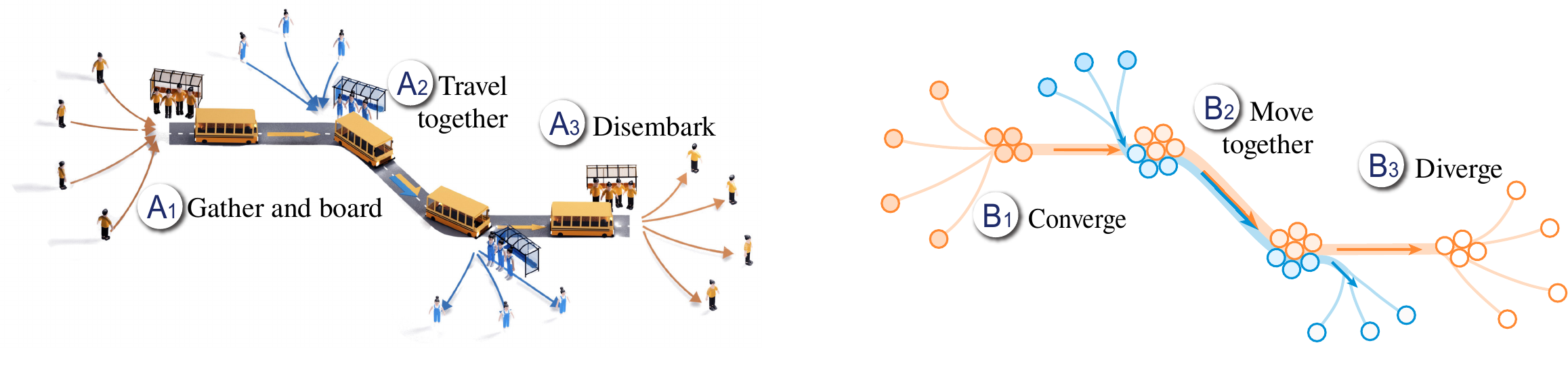}
  \put(-510,3){(a) Passengers gather and board the bus, travel together, and disembark}
  \put(-200,3){(b) Objects converge, move together, and diverge}
  \caption{Illustration of our analogy.
  }
  \Description{An illustration of our analogy.}
  \label{fig:teaser}
\end{figure*}

\section{Problem Formulation}
In this section, we introduce the problem formulation, including the bus route analogy and two sub-problems derived from this analogy.

\subsection{The Bus Route Analogy}

We illustrate the objects' movements in animation using the real-world bus route analogy, where passengers travel along different bus routes to reach their destinations.
As shown in Fig.~\ref{fig:teaser}(a), 
passengers gather at bus stops and board the same bus (A$_1$).
They then travel together along shared routes (A$_2$).
Eventually, they disembark at designated stops when approaching their destinations or transferring to other routes (A$_3$).
As such, we apply this analogy to guide the design of our animation.
As shown in Fig.~\ref{fig:teaser}(b), groups of objects with similar movement trajectories converge at local hotspots (B$_1$), analogous to bus stops, and then move together along the shared animation paths (B$_2$), much like passengers on the same bus.
As the animation progresses, these objects may diverge to reach their destinations separately (B$_3$).

Based on this analogy, we design our animation, RouteFlow, to capture both the global trend and local hotspots.
By grouping objects with similar trajectories and moving them along shared animation paths, we reveal the global trend, just as the bus routes that passengers travel along. 
Meanwhile, objects converge or diverge at specific local hotspots, similar to passengers boarding and disembarking at bus stops.
This allows us to simplify complex and cluttered trajectories in animation while ensuring that critical convergence and divergence points are preserved. 

Our animation leverages the Gestalt principles of \textit{Common Fate} and \textit{Proximity}~\cite{todorovic2008gestalt, wagemans2012century} to shape the perception of grouping. 
The \textit{Common Fate} principle states that visual elements moving together are perceived as a group~\cite{chalbi2020common}.
Accordingly, objects moving together along the same animation path are interpreted as a cohesive group.
The \textit{Proximity} principle states that visual elements close to one another are perceived as part of the same group~\cite{wagemans2012century}.
In this case, we position objects with similar trajectories in close proximity, simulating passengers on the same bus. 

To create the animation, we should generate the animation paths in a way that is similar to planning bus routes. 
Furthermore, since multiple objects often move simultaneously along the same animation path, we should minimize occlusion in animation, ensuring that each object has its own position, like passengers having individual seats on a bus.
In this process, the seat allocation depends on the bus routes, as the bus routes determine which passengers are on the bus and where they board and disembark.
This dependency naturally lends itself to sequential optimization~\cite{aubry2018sequential}.
In sequential optimization, the overall problem is decomposed into smaller, manageable sub-problems that are solved in sequence.
The solution to each sub-problem then informs and serves as the input for the subsequent one, ensuring a cohesive and efficient resolution of the entire problem.
Accordingly, we decompose the problem into two sub-problems: trajectory-driven path generation (bus routing) and object layout generation (seat allocation).
Next, we detail these two sub-problems and their respective optimization goals.

\subsection{Trajectory-Driven Path Generation}
\label{sec:formulation1}

In the context of the bus routing problem, there are two main optimization goals: efficiency and effectiveness~\cite{li2002school}. 
Efficiency involves minimizing operational costs, such as reducing the total length of the bus routes. 
One of the most effective strategies is route aggregation.
This strategy encourages passengers to share the same route as much as possible during their journeys.
The key is to identify groups of passengers with similar trajectories and then design routes that accommodate these shared trajectories.
Likewise, in animation, we should group objects with similar trajectories and share their animation paths to reduce the total path length.
This consolidates similar movements, allowing users to easily perceive the global trend in animation.
However, overemphasizing efficiency can lead to excessive route aggregation, forcing some passengers to deviate far from their intended trajectories and causing significant detours. 
On the other hand, effectiveness focuses on meeting passengers' travel demands by ensuring they can successfully reach their destination without excessive detours.
This requires minimizing the deviation between the aggregated route and each passenger's intended trajectory.
One solution is to set up proper bus stops in high-demand locations to satisfy more passengers' demands.  
In animation, the resulting animation paths should align closely with the input trajectories of the objects and thus better reveal critical local hotspots where objects converge or diverge.

As such, we derive two primary optimization goals for trajectory-driven path generation: 
\begin{itemize}
    \item \textbf{Minimize the total animation path length} by aggregating animation paths for groups of objects with similar trajectories. 
    \item \textbf{Minimize the deviation from the input trajectories} by constraining the distance between the input trajectories to their aggregated paths.  
\end{itemize}

\setcounter{figure}{2}
\begin{figure*}[!ht]
  \centering
  \includegraphics[width=0.9\linewidth]{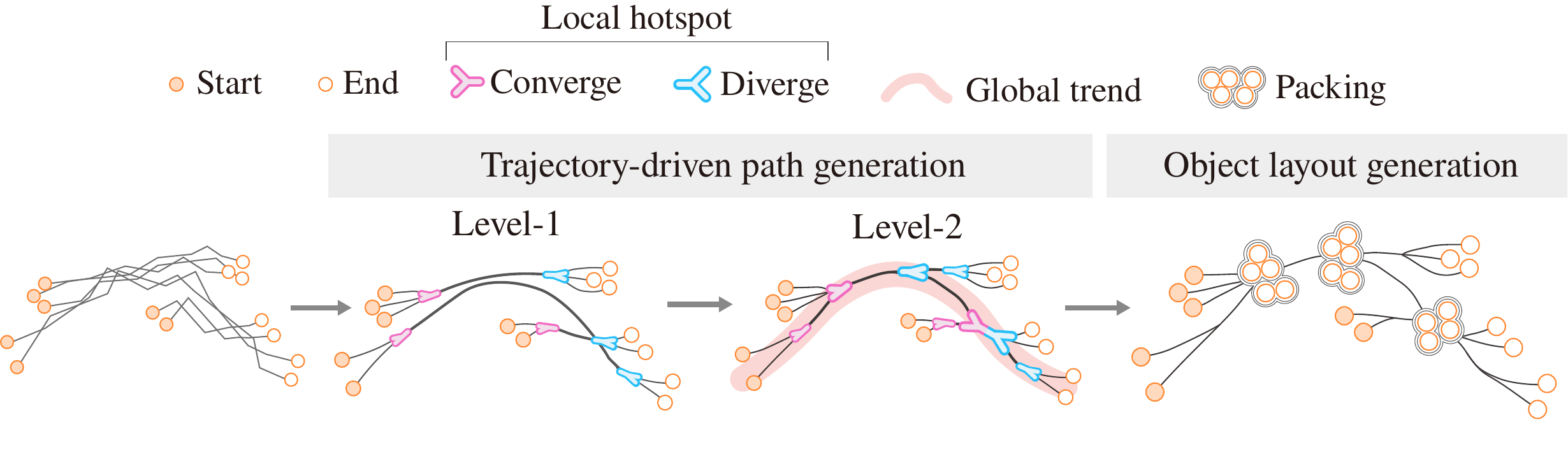}
  \put(-447,3){(a) Trajectory data}
  \put(-344,3){(b) Bottom-up hierarchical edge bundling algorithm}
  \put(-139,3){(c) Incremental circle packing algorithm}
  \caption{The pipeline of our method.
  }
  \label{fig:pipeline}
  \Description{The pipeline of our method.}
  \vspace{-3mm}
\end{figure*}

\subsection{Object Layout Generation}
\label{sec:formulation2}
There are two main optimization goals when allocating seats: maximize capacity and avoid overcrowding.
The first optimization goal is to maximize capacity.
Similar to buses efficiently filling seats, the object layout should be designed to reduce empty space to enhance compactness.
To achieve this, we strive to position similar objects close together to reduce gaps between them.  
This not only optimizes space utilization but also fosters a sense of group cohesion among closely placed objects, aligning with the \textit{Proximity} principle.
The second optimization goal is to avoid overcrowding, which can be addressed through three strategies.
First, when passengers are on the same bus, each should have their own seat to avoid interfering with others.
Second, co-travelers who board or disembark together should sit close to maintain group cohesion and avoid mixing with the crowd.
Third, passengers who disembark first should sit closest to the exit (the principle of ``first out, closest to the exit''), facilitating a smoother queueing process and mitigating potential overcrowding during disembarkation.
Correspondingly, in our animation: 1) objects moving along the same path should remain visible and not overlap; 2) objects that converge or diverge together should be grouped closely to avoid mixing with other groups;
and 
3) objects should be placed based on their disembarking order and positions.
Based on the analysis above, this problem involves two primary optimization goals: 
\begin{itemize}
    \item \textbf{Maximizing compactness} by reducing empty space in the layout.
    \item \textbf{Minimizing occlusion} by 1) applying the non-overlapping constraint within a group of objects moving together, 2) keeping objects that converge or diverge together as a group, and 3) following the principle of ``first out, closest to the exit.''
\end{itemize}



\section{Method}
Fig.~\ref{fig:pipeline} shows the pipeline of our method.
Given trajectory data as input, it consists of two modules: \textbf{trajectory-driven path generation} and \textbf{object layout generation}.

\subsection{Trajectory-Driven Path Generation}
An edge bundling algorithm can effectively minimize both the total path length and the deviation from the input trajectories in generating the animation paths.
However, aggregating all the trajectories simultaneously presents two issues.
First, it fails to identify local hotspots at multiple levels of granularity, which are pervasive in real-world applications~\cite{martino2019granular}.
Second, the real-time animated transitions require high scalability of the algorithm.
To address these issues, we develop a bottom-up hierarchical edge bundling algorithm that progressively bundles similar trajectories, level by level.
As shown in Fig~\ref{fig:pipeline}(b), it captures local hotspots across multiple levels of granularity while revealing the global trend.
At each level, we adopt a force-directed strategy~\cite{holten2009force, selassie2011divided} to bundle the edges.
The core of our algorithm lies in the design of the forces that drive the bundling process, along with a bottom-up bundling that progressively bundles trajectories.

\setcounter{figure}{3}
\begin{figure}[H]
  \centering
  \includegraphics[width=0.9\linewidth]{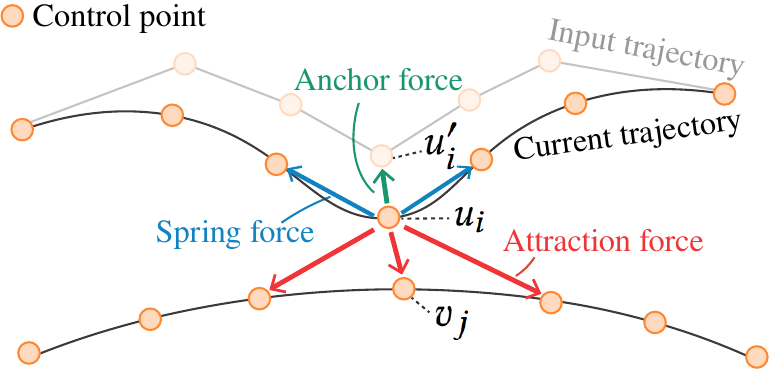}
  \caption{Illustration of three types of forces in our algorithm.}
  \label{fig:forces}
  \Description{Illustration of forces in our algorithm.}
\end{figure}

\setcounter{figure}{4}
\begin{figure*}[!hb]
 \centering
  \includegraphics[width=0.75\linewidth]{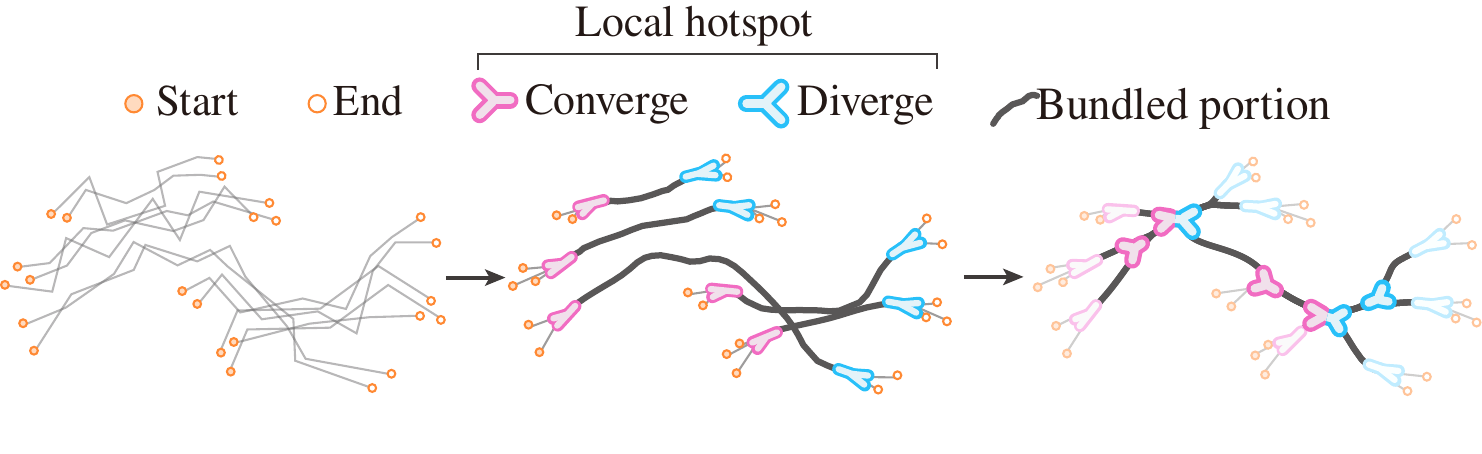}
  \put(-357,3){(a) Trajectory data}
  \put(-214,3){(b) Level-1}
  \put(-79,3){(c) Level-2}
  
  \caption{
  Illustration of our bottom-up hierarchical edge bundling algorithm.
  }
  \label{fig:global}
  \Description{Illustration of our bottom-up hierarchical edge bundling algorithm.
  }
\end{figure*}

\subsubsection{Force Design}
Existing force-directed edge bundling algorithms model trajectories as a series of control points and apply forces to adjust their positions~\cite{holten2009force,selassie2011divided}.
They typically adopt two types of forces: attraction force and spring force.
However, they often fail to preserve local hotspots because these forces ignore the original positions of these input trajectories.
To address this issue, we introduce a new force, the anchor force, to reduce deviation from the input trajectories.
Fig.~\ref{fig:forces} illustrates how our algorithm incorporates these three types of forces.
Given the trajectory set $S$ and a pair of trajectories $u$ and $v$, the three types of forces are defined as follows:

\begin{itemize}
    \item \textbf{Attraction force} ($F_{att}$) is applied between control points on different trajectories to draw them closer together.
    This force bundles similar trajectories.
    According to Selassie~\etal~\cite{selassie2011divided}, $F_{att}$ is defined as:
    \begin{equation}
        \begin{aligned}
            F_{att}(u_i,v_j)=\frac{\eta(v_j-u_i)}{C_v(\eta^2+||u_i-v_j||^2)^2},
        \end{aligned}
    \end{equation}
    where $u_i$ and $v_j$ represent the $i$-th and $j$-th control points on these trajectories, and $||u_i - v_j||$ denotes the Euclidean distance between them.
    The weighting parameter $\eta$ controls the rate at which the force diminishes with increasing distance.
    A larger $\eta$ causes $F_{att}$ to decrease slower, thereby extending its influence range.
    $C_v$ denotes the number of control points on trajectory $v$.
    \item \textbf{Spring force} ($F_{spr}$) is applied between adjacent control points on the same trajectory.
    This force promotes uniform distribution of control points along the trajectory and avoids highly curved trajectories.
    According to Holten~\etal~\cite{holten2009force}, $F_{spr}$ is defined as:
    \begin{equation}
        \begin{aligned}
            F_{spr}(u_i) = C_u(u_{i+1}+u_{i-1}-2u_i),
        \end{aligned}
    \end{equation}
    where $C_u$ is the number of control points on trajectory $u$.
    \item \textbf{Anchor force} ($F_{anc}$) is applied to each control point, pulling it back toward its position in the input trajectories.
    This force prevents the current trajectories from deviating too far from the input trajectories. $F_{anc}$ is defined as:
    \begin{equation}
        \begin{aligned}
            F_{anc}(u_i)=||u_i'-u_i||^2\cdot \frac{u_i'-u_i}{||u_i'-u_i||},
        \end{aligned}
    \end{equation}
    where $u_i'$ denotes the original position of $u_i$, and $\frac{u_i'-u_i}{||u_i'-u_i||}$ is a unit vector indicating the direction of the force.
\end{itemize}

Based on the above force analysis, the resultant force on the $i$-th control point of trajectory $u$ is calculated as:
\begin{equation}
    \begin{aligned}
        F(u_i) = (\sum_{v\in \Gamma_{u}}\sum^{C_v}_{j=1}F_{att}(u_i,v_j))+\alpha F_{spr}(u_i) +\beta F_{anc}(u_i).
    \end{aligned}
\end{equation}
Here, $\Gamma_{u}$ denotes the set of top-$k$ similar trajectories of $u$.
The parameters $\alpha$ and $\beta$ balance the three types of forces. In our implementation, they are determined as 5 and 1 through a grid search.

\subsubsection{Bottom-Up Hierarchical Edge Bundling}
Progressively bundling similar trajectories at each level of granularity involves two key aspects.
The first is how to select the most similar trajectories to consider when applying forces at each level.
Existing edge bundling algorithms assess edge similarities through compatibility metrics, which consider factors such as topology~\cite{selassie2011divided} and importance~\cite{Quan2012tgieb} but often fail to capture trajectory similarities.
To better capture trajectory similarities, we design our compatibility metric based on dynamic time warping (DTW)~\cite{muller2007dtw}, a widely accepted metric for assessing trajectory similarity~\cite{zheng2015trajectory}.
DTW calculates the distance between two trajectories by finding the optimal alignment between points on them, thereby capturing the overall similarity between the entire trajectories~\cite{muller2007dtw}.
Given two trajectories ($u$, $v$) and their DTW distance (DTW$(u,v)$), the compatibility between $u$ and $v$ is defined as:
\begin{equation}
    \begin{aligned}
    \text{compatibility}(u,v)=1 - \text{norm}(\text{DTW}(u,v)),
    \end{aligned}
\end{equation}
where norm$(\cdot)$ denotes the min-max normalization to scale the distance into the range [0,1].
To reduce computational complexity, we only consider the top-$k$ similar trajectories according to the compatibility metrics.
In our implementation, $k$ is a user-specified parameter that is set as five by default.

The second is how to hierarchically bundle these trajectories.
At each level, similar trajectories are bundled, revealing local hotspots where they converge and diverge.
To identify local hotspots across multiple levels of granularity, it is crucial to preserve the convergence and divergence identified at lower levels.
Therefore, we determine the bundled portions of trajectories at each level by measuring the distances between control points, as shown in Fig.~\ref{fig:global}.
Each bundled portion will be merged into a single trajectory, which serves as input for the next level, where they are further bundled.
Throughout this process, the identified local hotspots remain unchanged, as they are excluded from the bundled portions and unaffected by further bundling.
At each level, the three types of forces are applied.
When moving to the next level, the attraction force is increased tenfold to adapt to the sparser distribution of trajectories.

To evaluate the effectiveness of our algorithm, we compare it with two representative edge bundling algorithms, divided edge bundling (DEB)~\cite{selassie2011divided} and multilevel agglomerative edge bundling (MAEB)~\cite{gansner2011MINGLE}. 
We assess these methods based on their efficiency in reducing total path length and deviation from the original paths.
The results show that our algorithm achieves the lowest deviation and performs comparably with the baseline algorithms in terms of ink ratio.
Details can be found in Appendix~\ref{sec:appendixA}.

\setcounter{figure}{5}
\begin{figure}[t]
  \centering
  \setlength{\abovecaptionskip}{1mm}
  \includegraphics[width=0.9\linewidth]{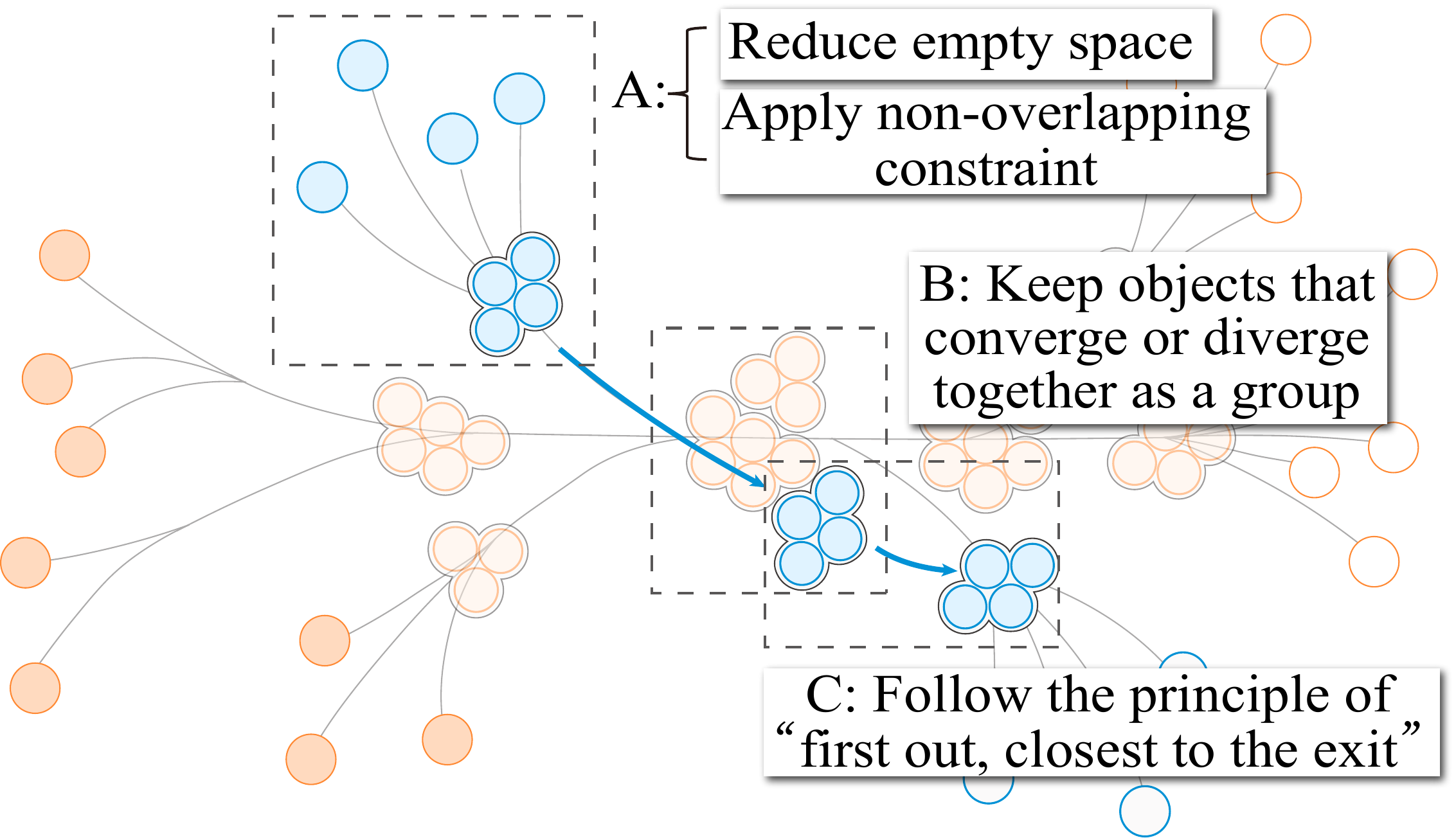}
  \caption{Illustration of the goals in our object layout generation.}
  \label{fig:packinggoal}
  \Description{Illustration of goals in our object layout generation.}
\end{figure}

\setcounter{figure}{6}
\begin{figure*}[hb]
  \centering
  \includegraphics[width=0.8\linewidth]{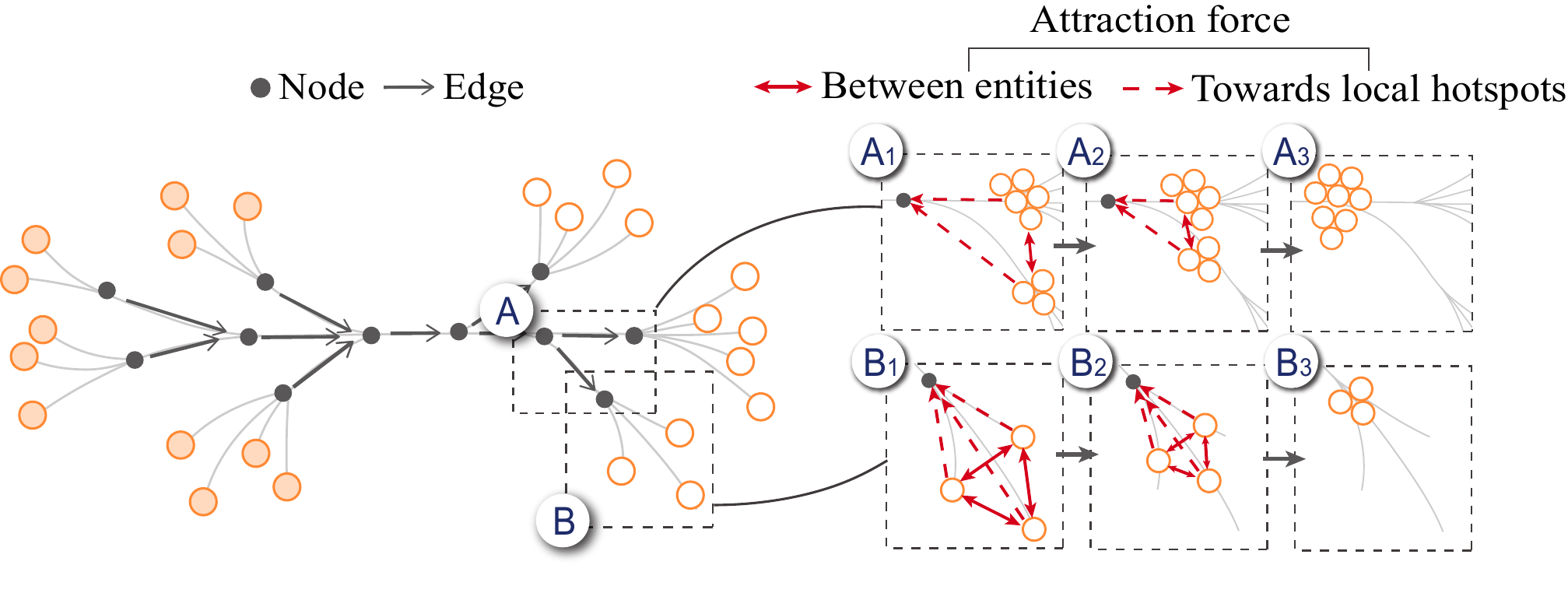}
  \put(-408,3){(a) Build a DAG to determine the order of local hotspots}
  \put(-184,3){(b) Generate the layout at each local hotspot}
  \caption{
  Illustration of our incremental circle packing algorithm.
  }
  \label{fig:packingpipeline}
  \Description{Illustration of our incremental circle packing algorithm.}
\end{figure*}

\subsection{Object Layout Generation}

The optimization goals described in Sec.~\ref{sec:formulation2} are achieved in three ways.
First, to reduce the empty space and satisfy the non-overlapping constraint for a group of objects (Fig.~\ref{fig:packinggoal}A), we use a circle packing algorithm~\cite{yuan2023visual} to generate the object layout.
Second, to keep objects that converge or diverge together as a group (Fig.~\ref{fig:packinggoal}B), an incremental circle packing algorithm is developed.
Third, to follow the principle of ``first out, closest to the exit"(Fig.~\ref{fig:packinggoal}C), we place objects based on their disembarking order and positions. 

Just as passengers only adjust their seats when boarding or disembarking along the bus route, we update the layout incrementally only at the local hotspots.
To achieve this, we first determine the order of local hotspots for layout generation by constructing a directed acyclic graph (DAG, Fig.~\ref{fig:packingpipeline}(a)) and then incrementally generate the layout at each local hotspot (Fig.~\ref{fig:packingpipeline}(b)).
In the DAG, nodes represent local hotspots, and directed edges indicate the movements of objects between these local hotspots.
We perform a reverse topological sort on the graph to generate the order of local hotspots.
Then, we incrementally generate the layout at local hotspots according to their order.
The basic idea of generating the layout at each local hotspot is to generate a layout for each newly arriving group and then pack these new layouts with those of previous groups.
As shown in Fig.~\ref{fig:packingpipeline}, when packing the objects at a given local hotspot A, they are placed to preserve their relative positions.
This prevents occlusion during disembarking.
Inspired by G\"ortler~\etal~\cite{gortler2018bubble}, we adopt a force-directed algorithm and apply two types of attraction forces (Fig.~\ref{fig:packingpipeline}(b)).
The first force moves all objects/groups toward the current local hotspot,
while the second attracts neighboring objects/groups together.
To avoid occlusion, we model objects/groups as rigid bodies and use the Box2D engine~\cite{catto2010box2d} for implementation.

\subsection{Implementation}
After generating the animation paths and object layout at all the local hotspots, we use an interpolation-based method to render smooth animations.
This method synchronizes the movements based on the timing of the local hotspots and the objects' start and end points. 
To simplify, we refer to these collectively as ``point.''
We ensure that 1) groups of objects that converge or diverge together arrive at or leave the local hotspots at the same time and 2) excessively fast speeds are avoided.
As shown in Fig.~\ref{fig:scanline}, we use a scan line that moves through all points, assigning their timing as when they intersect with the scan line.
The movement direction of the scan line is determined by the vector formed between the average start and end positions of all animation paths.
However, paths that form a large angle with the scan line's movement direction can result in excessively high speeds of objects.
To address this, we iteratively adjust the timing of points until the maximum speed is less than twice the minimum speed.
If we detect excessively high speed between two connected points, we adjust the timing by either delaying the latter point or advancing the former at random.
We interpolate between points to render the animation after completing the iterative adjustment.
To further enhance smoothness, we apply the slow-in, slow-out technique~\cite{dragicevic2011distortion}.
\setcounter{figure}{7}
\begin{figure}[t]
  \centering
  \includegraphics[width=0.8\linewidth]{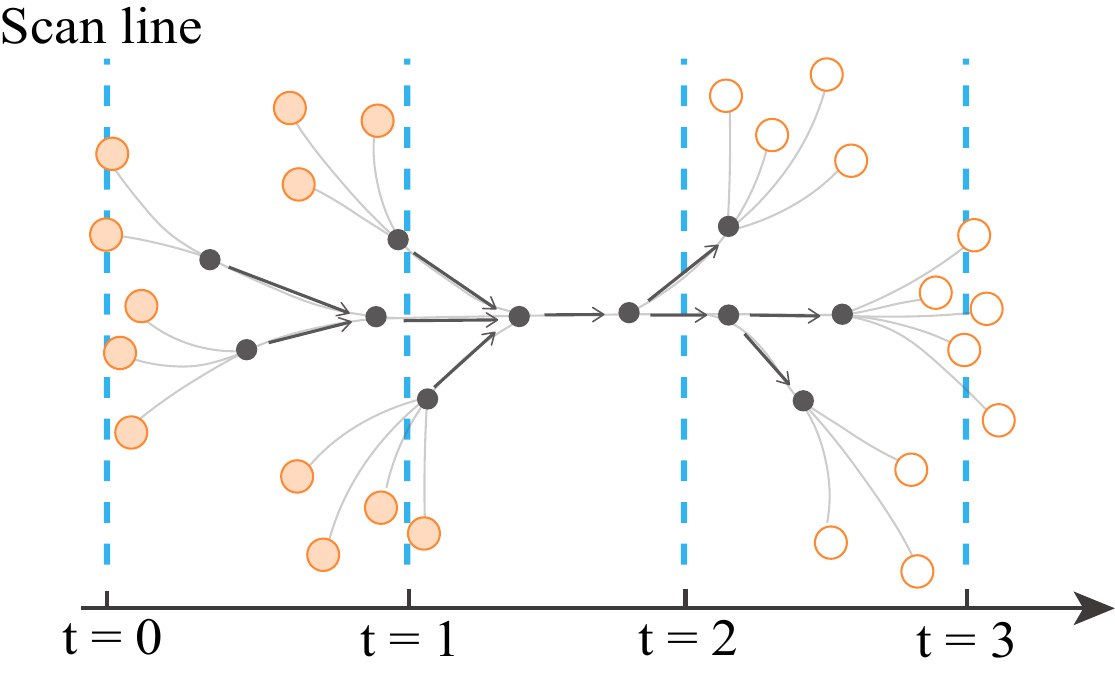}
  \caption{
  Illustration of the scan line moving through all the local hotspots and the objects' start and end points.
  }
  \label{fig:scanline}
  \Description{Illustration of our object layout.}
\end{figure}




\begin{table*}[b]
  \centering
  \setlength{\abovecaptionskip}{1.2mm}
  \caption{Comparison between different methods, including the focus+context grouping method (F+C), the vector-field-based method (VF), and RouteFlow. For all four metrics, lower values are better.}
  \label{tab:metric_result} 
  \resizebox{\textwidth}{!}{
    \begin{tabular}{c|ccccccccccccccccccccc}
       \toprule
        \multirow{2}{*}{\textbf{Dataset}}
        & \multicolumn{3}{c}{Overall occlusion}
        & \multicolumn{3}{c}{Within-group occlusion}
        & \multicolumn{3}{c}{Deformation}
        & \multicolumn{3}{c}{Dispersion}\\
        \cmidrule(lr){2-4} \cmidrule(lr){5-7} \cmidrule(lr){8-10} \cmidrule(lr){11-13}
        & {F+C} & {VF} & {RouteFlow}
        & {F+C} & {VF} & {RouteFlow}
        & {F+C} & {VF} & {RouteFlow}
        & {F+C} & {VF} & {RouteFlow} \\
        \midrule

        Taxi~\cite{yuan2011drive, yuan2010tdrive} &
        0.00123 &	0.00180 &	\textbf{0.00048} &
        0.01963 &	0.00626 &	\textbf{0.00606} &
        0.00081 &	0.00107 &	\textbf{0.00062} &
        0.05948 &	0.06766 &	\textbf{0.02606} \\

        BirdMap~\cite{BirdMap} & 
        0.00277 &	0.00958 &	\textbf{0.00213} &
        0.02773 &	0.02607 &	\textbf{0.00904} &
        0.00152 &	0.00160 &	\textbf{0.00064} &
        0.13762 &	0.12494 &	\textbf{0.03258} \\

        Railway~\cite{railway} &
        0.00229 &	0.01351 &	\textbf{0.00142} &
        0.11960 &	0.10420 &	\textbf{0.02875} &
        0.00128 &	0.00140 &	\textbf{0.00063} &
        0.10629 &	0.10950 &	\textbf{0.03090} \\

        MEIBook~\cite{mei} & 
        0.00132 &	0.00566 &	\textbf{0.00066} &
        0.01362 &	0.01196 &	\textbf{0.00675} &
        0.00079 &	0.00105 &	\textbf{0.00064} &
        0.06699 &	0.06817 &	\textbf{0.02430} \\

        OpenSkyAirline~\cite{opensky} &
        0.00076	&   0.00394	&   \textbf{0.00055} &
        0.01934	&   0.01814 &  	\textbf{0.00571} &
        0.00082	&   0.00117	&   \textbf{0.00061} &
        0.08047	&   0.08216	&   \textbf{0.03007} \\

        US Migration~\cite{holten2009force} & 
        0.00115 &	0.00408 &	\textbf{0.00072} &
        0.01107 &	0.01027 &	\textbf{0.00295} &
        0.00097 &	0.00147 &	\textbf{0.00066} &
        0.13015 &	0.13345 &	\textbf{0.03802} \\

        DanishAIS~\cite{DanishAIS} &
        0.00047 &	0.00120 &	\textbf{0.00012} &
        0.01435 &	0.00460 &	\textbf{0.00266} &
        0.00122 &	0.00089 &	\textbf{0.00087} &
        0.11688 &	0.12816 &	\textbf{0.08611} \\

        \midrule
        
        \textbf{Average} & 
        0.00148 &	0.00561 &	\textbf{0.00088} &
        0.03097 &	0.02469 &	\textbf{0.00842} &
        0.00122 &	0.00141 &	\textbf{0.00065} &
        0.10627 &	0.10858 &	\textbf{0.03727} \\

        \bottomrule
    \end{tabular}
    }
\end{table*}


\section{Evaluation}
To demonstrate the effectiveness of RouteFlow, we conducted a quantitative experiment and a user study.

\subsection{Quantitative Evaluation}
\label{sec:quantitative}

\noindent\textbf{Datasets}.
The quantitative evaluation was conducted on seven datasets from real-world applications: Taxi~\cite{yuan2011drive,yuan2010tdrive}, BirdMap~\cite{BirdMap}, Railway~\cite{railway},  MEIBook~\cite{mei}, OpenSkyAirline~\cite{opensky}, US Migration~\cite{holten2009force}, DanishAIS~\cite{DanishAIS}.
These datasets were collected from three common areas in trajectory data analysis, including transportation, sociology, and ecology.
We preprocessed the raw data through several steps, including noise filtering, trajectory compression, and merging of redundant trajectories.

\noindent\textbf{Baseline methods}.
We selected two state-of-the-art animated transition methods for comparison.
The first method, the focus+context grouping method, simultaneously facilitates tracking objects' movements and identifying the global trend by breaking down transitions into groups~\cite{zheng2018focus+}.
We used the default parameters reported in the paper.
The second method, the vector-field-based method, is the state-of-the-art method in terms of tracking objects' movements by utilizing vector fields to generate smooth, non-linear paths~\cite{wang2017vector}.
As the original paper did not provide specific parameter settings, we performed a grid search to find the optimal parameters.
Moreover, since the vector-field-based method requires predefined groups, we used the grouping results from the focus+context grouping method for consistency.

\setcounter{figure}{8}
\begin{figure*}[b]
  \centering
  \includegraphics[width=0.9\linewidth]{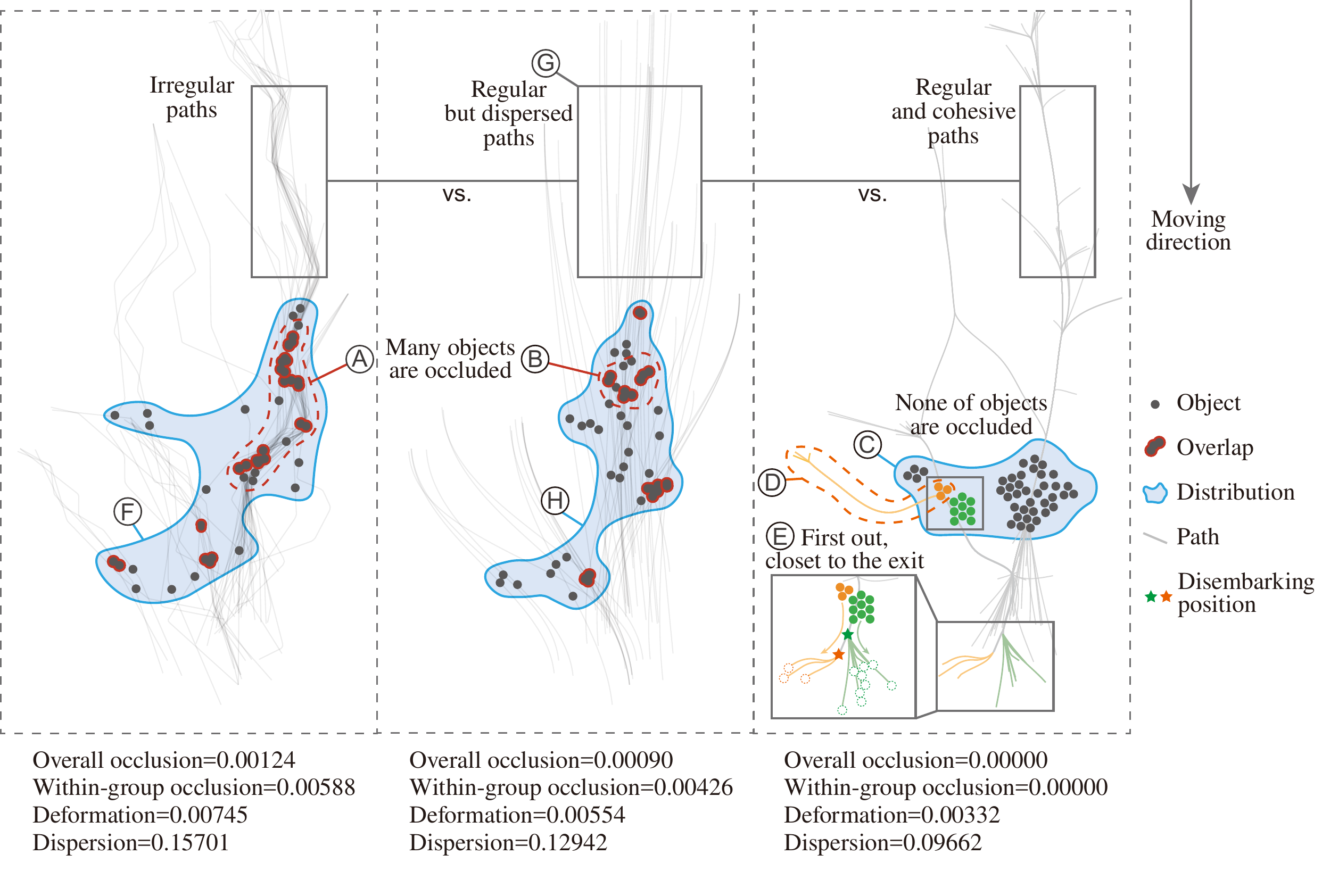}
  \put(-442,3){(a) Focus+context grouping}
  \put(-317,3){(b) Vector-field-based method}
  \put(-158,3){(c) RouteFlow}
  \caption{Object positions at a specific frame in the animations generated by three methods on the BirdMap dataset. Here, overlaps are highlighted as red strokes, and the distributions of objects are shown as blue contours. 
  The metric values for these frames are displayed below each sub-figure.
  The detailed analysis of these values is provided in Appendix~\ref{sec:appendixB}.}
  \label{fig:quan-result}
  \Description{User study results in three tasks, including Friedman tests and pairwise Wilcoxon signed-rank tests on three methods.}
\end{figure*}

\noindent\textbf{Evaluation criteria}. 
Previous studies classified the metrics into three types: occlusion, deformation, and dispersion~\cite{chevalier2014not, dragicevic2011distortion, du2015trajectory, wang2017vector}.
We adopted the metrics summarized by Wang~\etal~\cite{wang2017vector} as they are tailored for objects with groups.

We denote all the frames in the animation as $T$, a particular frame as $t$, and
all the objects as $P$.

\myparagraph{Occlusion} measures the overlap between objects.
This metric is useful for evaluating the capability of an animation in facilitating the tracking of objects' movements.
High occlusion reduces the visibility and distinguishability of moving objects, making these objects harder to distinguish and track~\cite{chevalier2014not,dragicevic2011distortion}.

Specifically, \textit{overall occlusion} (occlusion$_o$) measures the overlap between all objects during the entire animation.
\begin{equation}
\text{occlusion}_o(T) = \notag \frac{1}{|T|} \sum_{t \in T} \frac{\sum_{p,q \in P, p\neq q} \text{overlap}(p, q, t)}{|P|(|P| - 1)},
\end{equation}
where $\text{overlap}(p, q, t)$ is an indicator function with value 1 if objects $p$ and $q$ overlap at frame $t$, and 0 otherwise.

\textit{Within-group occlusion} (occlusion$_w$) measures the overlap between objects in the same group.
\begin{equation}
\begin{aligned}
\text{occlusion}_w(T) = \notag \frac{1}{K}\sum_{i=1}^{K}\frac{1}{|T_{G_i}|} \sum_{t \in T_{G_i}} \frac{\sum_{p,q \in G_i, p\neq q} \text{overlap}(p, q, t)}{|G_i|(|G_i|-1)},
\end{aligned}
\end{equation}
where $K$ is the number of groups, $G_i$ is the set of objects in the $i$-th group, and $T_{G_i}$ is the frames of the group $G_{i}$.

\myparagraph{Deformation} measures the changes in distance between objects within the same group across consecutive time frames.
Lower deformation indicates that the relative object positions within the group remain stable, making it easier to track the objects of interest.

{\setlength{\abovedisplayskip}{0.7mm}\setlength{\belowdisplayskip}{2mm}
\begin{equation}
\begin{aligned}
& \text{deformation}(T) =  \notag \\
& \frac{1}{K} \sum_{i=1}^{K} \frac{1}{|T_{G_i}|} 
 \sum_{t \in T_{G_i}, t > 0}
 \frac{\sum_{p, q \in G, p \neq q} |\text{dist}(p, q, t) - \text{dist}(p, q, t - 1)|}{|G_i| (|G_i| - 1)}.
\end{aligned}
\end{equation}}
$dist(p, q, t)$ is the distance between objects $p$ and $q$ at frame $t$.

\myparagraph{Dispersion} measures how spread out objects in the same group are.
Lower dispersion indicates that the members of a group are moving more closely together, enhancing the perception of the group as a whole.
This facilitates more effective tracking of the group's collective movements, thereby enhancing the identification of the global trend.

{\setlength{\abovedisplayskip}{0.7mm}\setlength{\belowdisplayskip}{2mm}
\begin{equation}
\begin{aligned}
\text{dispersion}(T) = \notag \frac{1}{K}\sum_{i=1}^{K}\frac{1}{|T_{G_i}|} \sum_{t \in T_{G_i}} \frac{\sum_{p,q \in {G_i}, p\neq q} \text{dist}(p, q, t)}{|{G_i}|(|{G_i}|-1)}.
\end{aligned}
\end{equation}
}

The aforementioned three metrics focus on tracking objects' movements and identifying the global trend.
To the best of our knowledge, there is no existing metric that adequately measures the preservation of local hotspots in animation.
Additionally, the employed real-world datasets lack ground truth for local hotspots. 
As a result, we supplement the evaluation of preserving local hotspots with a user study using several synthetic datasets, which is described in Sec.~\ref{sec:userstudy}.

\noindent\textbf{Results}. 
Table~\ref{tab:metric_result} presents the comparison results between RouteFlow and the baseline methods.
RouteFlow performs the best on all datasets and all metrics.


\begin{table*}[t]
  \centering
  \setlength{\abovecaptionskip}{1mm}
  \caption{The result of running time on different modules.}
  \label{tab:time_result} 
  \begin{tabular}{c|ccccc}
    \toprule
    \multirow{2}{*}{\textbf{Dataset}}
    & \multicolumn{1}{c}{Attribute}
    & \multicolumn{3}{c}{Time cost (second)} \\
    \cmidrule(lr){2-2} \cmidrule(l){3-5}
    & {Size}
    & {Trajectory-driven path generation}
    & {Object layout generation} & {Total} \\
    \midrule
    Taxi~\cite{yuan2011drive, yuan2010tdrive} &
     73 & 0.06 & 0.03 & 0.09 \\
    BirdMap~\cite{BirdMap} &
     109 & 0.10 & 0.06 & 0.16 \\
    Railway~\cite{railway} &
     129 & 0.11 & 0.06 & 0.17 \\
    MEIBook~\cite{mei} &
     174 & 0.11 & 0.10 & 0.21 \\
    OpenSkyAirline~\cite{opensky} &
     187 & 0.14 & 0.11 & 0.25 \\
    US Migration~\cite{holten2009force} &
     258 & 0.23 & 0.22 & 0.45 \\
    DanishAIS~\cite{DanishAIS} &
     316 & 0.46 & 0.37 & 0.83 \\
    \bottomrule
  \end{tabular}
  \vspace{2mm}
\end{table*}


\setcounter{figure}{9}
\begin{figure*}[b]
  \centering
  \setlength{\abovecaptionskip}{0mm}
  \includegraphics[width=0.8\linewidth]{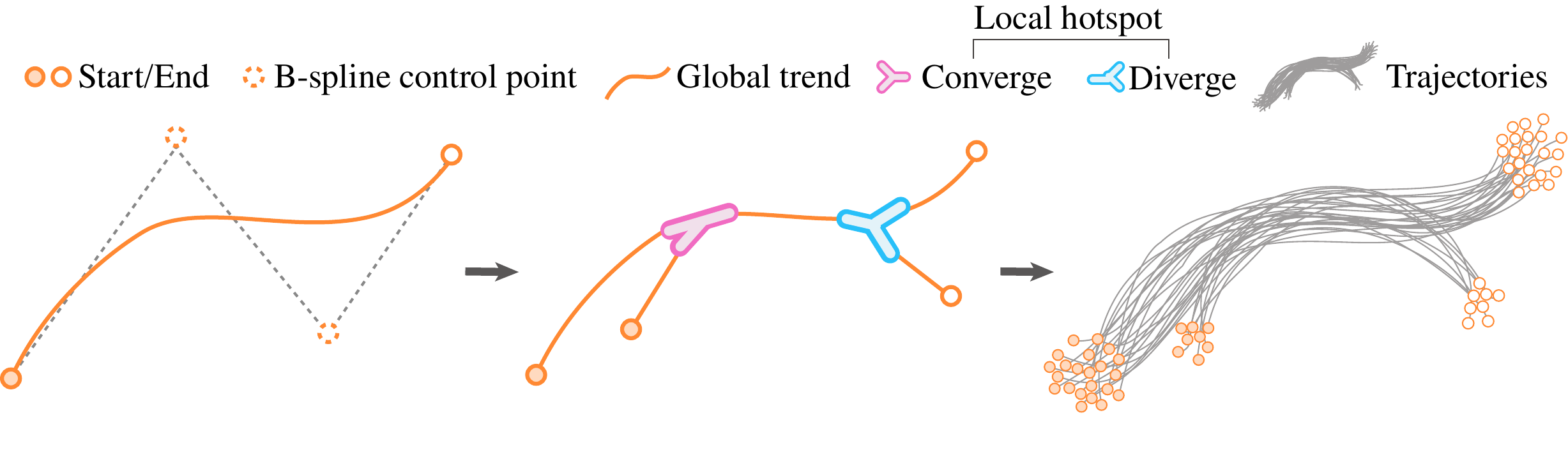}
  \put(-398,10){(a) Generate the global trend}
  \put(-261,10){(b) Determine local hotspots}
  \put(-108,10){(c) Create trajectories}
  \caption{The data generation pipeline.}
  \label{fig:datageneration}
  \Description{Illustration of forces in our edge bundling algorithm.}
\end{figure*}

\myparagraph{Occlusion}.
RouteFlow achieves lower overall occlusion and within-group occlusion scores compared to the two baseline methods.
This improvement is mainly due to differences in object layout. 
The baseline methods do not explicitly optimize the overlaps within groups (Fig.~\ref{fig:quan-result}A and Fig.~\ref{fig:quan-result}B), and in particular, the vector-field-based method may even introduce overlaps between groups as it moves all objects simultaneously. 
In contrast, our incremental circle packing algorithm reduces overlaps by employing three strategies: 1) applying a non-overlap constraint to minimize occlusion between objects (Fig.~\ref{fig:quan-result}C), 2) keeping objects that converge or diverge together as a group (Fig.~\ref{fig:quan-result}D), and 
3) following the principle of ``first out, closest to the exit'' (Fig.~\ref{fig:quan-result}E).

\myparagraph{Deformation}.
Compared to the two baseline methods, RouteFlow exhibits the least deformation.
The focus+context grouping method changes the layout of every frame as objects follow their input trajectories.
Similarly, the vector-field-based method causes unsynchronized movements due to varying velocity vectors among objects at different positions, leading to layout changes in subsequent frames. 
Conversely, RouteFlow incrementally updates the layout only at local hotspots, with the aim of balancing readability and stability in these refinements.
This balance greatly reduces deformation during animated transitions.

\myparagraph{Dispersion}.
RouteFlow achieves lower dispersion compared to the other methods.
The focus+context grouping method, which groups objects based solely on their start and end positions, records the highest dispersion. 
This method might group objects with different trajectories together due to their similar start and end positions, increasing dispersion (Fig.~\ref{fig:quan-result}F).
The vector-field-based method, which generates the animation paths of objects using a vector field, tends to disperse the input trajectories (Fig.~\ref{fig:quan-result}G), leading to a less compact layout (Fig.~\ref{fig:quan-result}H) and high dispersion.
In contrast, our incremental circle packing algorithm keeps the objects compact (Fig.~\ref{fig:quan-result}C), resulting in lower dispersion.

\noindent\textbf{Running Time}. 
\label{para:runningtime}
Table~\ref{tab:time_result} shows the average running times for each module of our animation method on real-world datasets, where object sizes vary from 73 to 316.
The performance tests were conducted on a Windows PC with an Intel i9-13900K CPU.
We averaged results over five trials to minimize randomness.
The average running time per dataset is within 1 second, which is fast enough for designing animated transitions.
The object layout generation module is the most time-consuming because it requires incremental generation for the layout of each local hotspot.
In contrast, the trajectory-driven path generation module is less demanding and achieves stable results in 300 iterations.

\subsection{User Study}
\label{sec:userstudy}
We conducted a user study to evaluate how effectively people use RouteFlow to track objects' movements and identify the global trend and local hotspots.
We formulated three hypotheses: participants perform more accurately with RouteFlow in tracking objects' movements (\textbf{H1}), identifying the global trend (\textbf{H2}), and locating local hotspots (\textbf{H3}) compared to two baseline methods, the focus+context grouping method and the vector-field-based method.

\newcommand{\track}{\textit{T1---Tracking objects' movements}}
\newcommand{\trend}{\textit{T2---Identifying the global trend}}
\newcommand{\hotspot}{\textit{T3---Locating local hotspots}}
\newcommand{\myquote}[1]{\textit{``#1''}}

\subsubsection{Study Setup}

\noindent\textbf{\\ Participants}. 
We recruited 15 participants (12 males and 3 females, denoted as P1-P15) from local universities.
They were graduate students majoring in computer science (12) and information design (3), aged from 22 to 32 years (\textit{mean} = 24.47, \textit{SD} = 2.53).
All of them reported to have normal vision and no color deficiencies.
Upon completion, each participant received a \$30 compensation, independent of their performance.

\noindent\textbf{Apparatus}. 
The user study was conducted on a personal computer equipped with a 27-inch display with a resolution of 3840 $\times$ 2160 pixels and a 60 Hz refresh rate.
Objects were presented as circles with a radius of 9 pixels (0.20 cm), filled in black color, following the previous practice~\cite{du2015trajectory,wang2017vector}.
The animation window measured 1250 $\times$ 1250 pixels (27.0 $\times$ 27.0 cm) with a white background.
Participants were seated at a distance of 40 cm from the display.

\noindent\textbf{Datasets}.
We used synthetic data for a controlled study setting instead of real-world data, which may lack ground truth for the global trend and local hotspots.
This follows the common practice~\cite{du2015trajectory,wang2017vector}. 
Fig.~\ref{fig:datageneration} shows our dataset generation process, involving three steps.
First, we generated a smooth global trend trajectory using B-splines.
Next, we determined local hotspots by sampling points along the global trajectory and randomly classifying them as converging or diverging points.
Finally, we created trajectories for objects by adding random perturbations to the global trend, avoiding overlap between start and end positions.
To achieve better diversity and a certain level of complexity, we finalized our design through several iterations.
In the final iteration, there were two types of global trend (one or two bends in the B-spline), and three types of local hotspot assignment (1 convergence + 1 divergence, 2 convergences + 1 divergence, and 1 convergence + 2 divergences).
Each dataset included 30 trajectories.
To control the experiment duration and keep participants focused, we generated datasets for each combination of trend type and local hotspot type and conducted two repetitions per combination, resulting in 12 datasets (2 types of the global trend $\times$ 3 types of local hotspot assignment $\times$ 2 repetitions).

\noindent\textbf{Task design}.
Our study consisted of three tasks, each iterated and refined through small-scale pilot studies.
For each task, participants were asked questions with four options (one correct, three incorrect) along with an additional option for ``I am not sure.''

\track{}:
Participants were required to track the movement of target objects to identify their end positions, and then select one answer from five options.
This task design referred to the previous practices~\cite{du2015trajectory, wang2017vector}. 
We set the number of target objects to three, all from the same group, to simplify the task. 
We generated the incorrect options by randomly replacing the correct targets with their nearest neighbors based on their end positions.\looseness=-1

\trend{}:
Participants were asked to observe the overall movement of all objects to identify the global trend, and then select one answer from five options.
Initially, we set the background to be fully white, whereas feedback from the pilot study indicated difficulty in observing and locating the movements. 
To alleviate this issue, the background canvas was divided into 8 $\times$ 8 grid, colored alternatively in white and grey. 
The three incorrect options were generated by adding random perturbations to the correct trend, ensuring that they passed through different grids to be distinguishable from the correct option.

\hotspot{}:
Participants were asked to identify the grids containing local hotspots.
Similar to \textit{T2}, to facilitate locating local hotspots, we employed a white and grey background canvas.
Each answer option included two marked grids: one for convergence and one for divergence.
Incorrect options were generated from the correct option by randomly replacing one correct grid with a neighboring grid.
To simplify the task, participants were allowed to click and mark grids that might assist them while viewing the animation and refer to these marks when answering.

\noindent\textbf{Study protocol}.
Participants started by signing consent and watching a tutorial video about the study procedure and tasks. 
We then provided three brief videos, each explaining a different animation method.
The study adopted a within-subjects design, requiring each participant to complete all three tasks using three different methods.\looseness=-1

For each task, we designed a practice session and a test session. 
The practice session familiarized participants with the interface and tasks, through six trials, two for each method. 
In each trial, participants initially saw all objects in grey points. 
Particularly, we highlighted the target objects in red for the tracking task. 
They then clicked to start the animation. 
All objects in the tracking task transitioned to grey and then to black within the first 0.5 seconds. In the other two tasks, the objects turned black directly. This allowed participants to recognize the target objects and prepare to follow their movements.
After the animation, participants clicked to start the question and could not review the animation anymore. 
In the practice session, we provided correct answers to help participants check their understanding and encouraged them to ask questions.

After completing the practice session and confirming that they fully understood the tasks and methods, participants advanced to the test session, which consisted of 36 trials (12 datasets $\times$ 3 methods). 
Unlike the practice session, correct answers were no longer provided during the test session.
To counterbalance the order of methods, we divided 15 participants into five groups, three for each group. 
Within each group, we used an expanded Latin square and applied a cyclic shift to the method order for each participant.
Additionally, to alleviate the learning effect, we randomly mirrored and rotated the datasets. 
Participants were allowed to take short breaks after each task or whenever they requested one.

In total, each participant finished 108 trials (3 tasks $\times$ 3 methods $\times$ 12 datasets), leading to 1,620 total trials (15 participants $\times$ 108 trials). 
After finishing each task, we assessed participants' workload and fatigue levels using NASA's Task Load Index~\cite{sandra2006nasa} and asked about their preferred methods and the reasons for their preferences. For all trials, we recorded participants' answers and completion times.
The entire study lasted 75-90 minutes.
Additional study details are provided in Appendix~\ref{sec:appendixC}, and results are provided in the supplemental material.

\setcounter{figure}{10}
\begin{figure*}[t]
    \vspace{5mm}
  \centering
  \includegraphics[width=\linewidth]{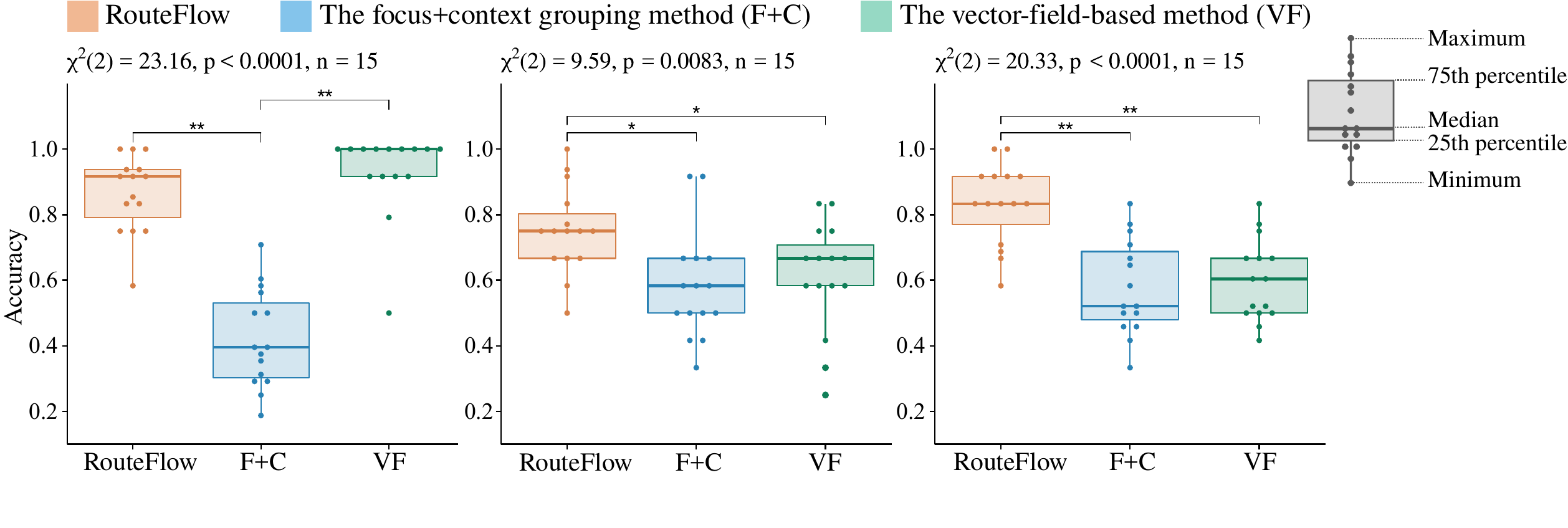}
  \put(-497,3){(a) T1 --- tracking objects' movements}
  \put(-353,3){(b) T2 --- identifying the global trend}
  \put(-206,3){(c) T3 --- locating local hotspots}
  \caption{
  User study results on three tasks. Here, * indicating \textit{p} < 0.05, ** indicating \textit{p} < 0.01.
  }
  \label{fig:user-study-result}
  \Description{User study results on three tasks, including Friedman tests and pairwise Wilcoxon signed-rank tests on three methods.}
  \vspace{2.9mm}
  
\end{figure*}

\subsubsection{Result Analysis} 
We analyzed three types of data: the accuracy of multi-choice questions, participants' subjective ratings for workload, and their stated preferences.  

\noindent\textbf{Accuracy}.
For each task, we computed participants' average accuracy across different methods.
As data is not normally distributed, we conducted Friedman tests for each task, followed by Wilcoxon signed-rank tests with Bonferroni correction for pairwise comparisons of different methods.
We report the statistical test results and the box plots in Fig.~\ref{fig:user-study-result}.
The Friedman test results indicate significant differences among the methods in all three tasks: tracking objects' movements ($\chi^2(2)=23.16, p < 0.0001$), identifying the global trend ($\chi^2(2)=9.59, p = 0.00083$), and locating local hotspots ($\chi^2(2)=20.33, p < 0.0001$).
In the subsequent analysis, we focus on the pairwise comparisons.

\track:
In pairwise comparisons, our RouteFlow method significantly outperforms the focus+context grouping method (average: 0.865 vs. 0.421, $p = 0.001$).
However, RouteFlow only shows a small difference compared to the vector-field-based method in average accuracy (0.865 vs. 0.930), and the paired test indicates no significant difference ($p = 0.275$).
Therefore, the results can \textbf{partially support H1}.
Participants using the focus+context grouping method show a significantly lowest accuracy.
This is because both our method and the vector-field-based method optimize the animation paths to reduce complexity, while the focus+context grouping method directly uses the input trajectories as its animation paths.
Correspondingly, the majority of participants (10 out of 15) reported that the animation paths in the focus+context grouping method were more complex and had greater occlusion compared to the other two methods.
This feedback aligns with our comparison results in Sec.~\ref{sec:quantitative}, where the focus+context grouping method achieves the highest within-group occlusion.
In addition, we asked participants about their different performances using RouteFlow and the vector-field-based method. 
P12 appreciated the grouping in RouteFlow for tracking, saying \myquote{The relative positions of objects are stable within the group. I can track one or two targets inside the group and use them as a reference to locate other targets.} 
However, we also received feedback expressing different opinions. For instance, P2 commented, \myquote{The objects in RouteFlow are grouped closely. I sometimes confuse the targets with other objects.} P2 preferred the vector-field-based method, noting \myquote{objects are relatively spread out. Even when occlusion occurs, it is easy to distinguish the targets by following their movements.}

\trend:
RouteFlow significantly outperforms the focus+context grouping method ($p = 0.021$) and the vector-field-based method ($p=0.028$), thereby \textbf{supporting H2}.
Participants described RouteFlow as having a sense of \myquote{unity} (P1, P2, P4, P10, and P11) and showing \myquote{a clear trend} (P1, P4, P5, P6, and P11).
Besides, they explained their performances using the two baseline methods related to the dispersion. 
P2 complemented that \myquote{Although [the focus+context grouping method] depicts a clear trend of each group, it requires memorizing and comparing the movements of several groups to get the global trend, which brings extra burden.}
P4 mentioned the vector-field-based method, saying \myquote{It is distracting to follow the objects that are dispersed in different locations but move synchronously.}
This feedback aligns with the quantitative result in Sec.~\ref{sec:quantitative}, where these baseline methods showed higher dispersion compared to RouteFlow.
Additionally, P1 commented that the animation design for groups of objects in RouteFlow can be a double-edged sword for capturing the global trend: \myquote{[It allows me] to infer the global trend based on groups' movements. However, it may be disrupted due to the convergence and divergence [of the groups].}
This further highlighted the importance of balancing the global trend and local hotspots.

\hotspot:
RouteFlow achieves an average accuracy of 0.832, which is significantly better than the focus+context grouping method ($p = 0.002$) and the vector-field-based method ($p=0.002$).
Therefore, the results can \textbf{support H3}.
Participants described the convergence and divergence of objects in RouteFlow as \myquote{clearly noticeable} (P1, P2, P12, P13, and P15) and \myquote{apparent} (P6, P8, and P10).
In contrast, the other two methods required extra cognitive effort.
Five participants (P2, P6, P11, P12, and P15) complained that animating different groups separately in the focus+context grouping method made it difficult to distinguish the locations where objects converge or diverge.
Although the vector-field-based method animated all objects together, users still found it difficult to identify local hotspots.
P11 commented from the spatial perspective, \myquote{Objects are dispersed. I cannot confidently tell the locations [of the local hotspots].}
P2 noticed \myquote{Objects do not pass through their shared positions simultaneously.}
This issue arises because, while the vector-field-based method animates objects concurrently, the local hotspot locations along their individual trajectories do not always align.
Thus, participants noted temporal discrepancies in the animation as objects passed through these local hotspots.\looseness=-1

\noindent\textbf{Workload}.
Fig.~\ref{fig:NASA} summarizes participants' workload and fatigue levels using NASA's Task Load Index~\cite{sandra2006nasa}, including mental demand, physical demand, temporal demand, effort, frustration, and performance.
We conducted Friedman tests for each dimension in each task, followed by Wilcoxon signed-rank tests with Bonferroni correction for pairwise comparison among different methods.
The Friedman test results show significant differences among the methods across all six dimensions for each task, except the effort dimension for \trend. 
Next, we focus on analyzing the pairwise comparison results. 

\setcounter{figure}{11}
\begin{figure*}[t]
  \centering
  \includegraphics[width=\linewidth]{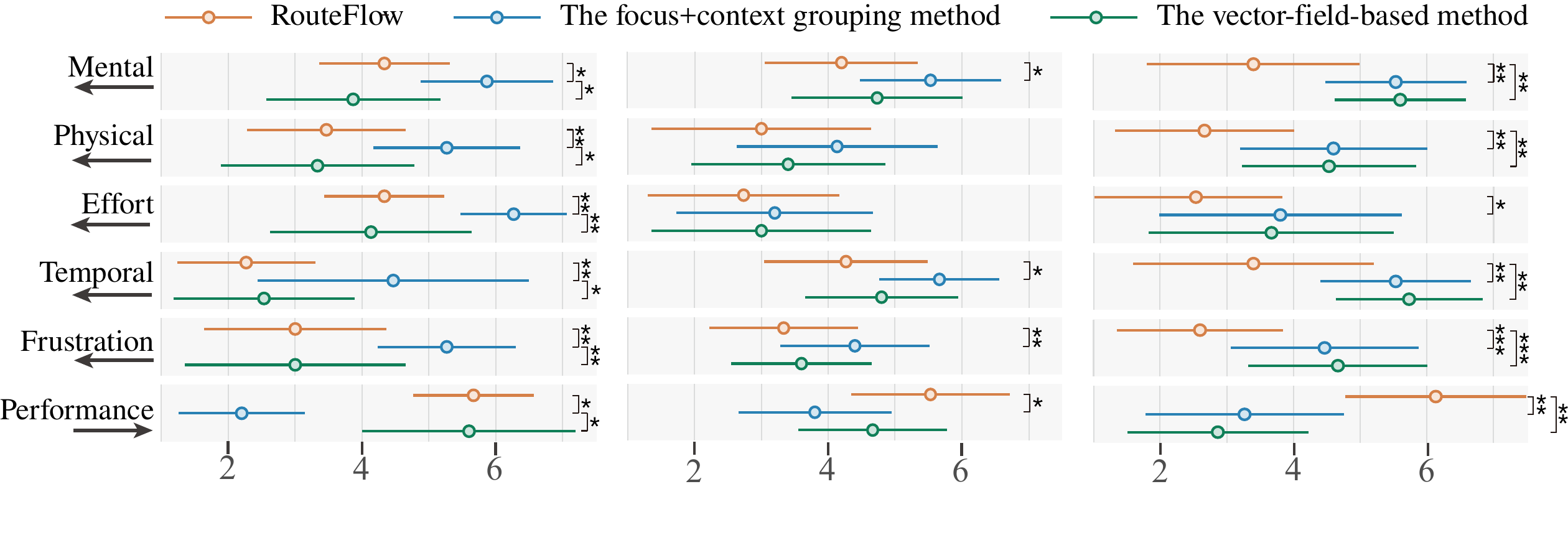}
  \put(-453,3){(a) T1 --- tracking objects' movements}
  \put(-303,3){(b) T2 --- identifying the global trend}
  \put(-142,3){(c) T3 --- locating local hotspots}
  \caption{Participants' workload and fatigue levels according to NASA's Task Load Index.
  Here, error bars show the 95\% confidence intervals.
  $\vcenter{\hbox{\includegraphics[height=0.012\textwidth]{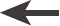}}}$ indicates that a smaller score is better, and vice versa.
   The results of pairwise Wilcoxon signed-rank tests are also shown, with * indicating \textit{p} < 0.05, ** indicating \textit{p} < 0.01, and *** indicating \textit{p} < 0.001.
  }
  \Description{Participants' workload and performance according to NASA's Task Load Index.}
  \label{fig:NASA}
\end{figure*}

\track: 
The Wilcoxon signed-rank tests indicate that both RouteFlow and the vector-field-based method perform better than the focus+context grouping method on all six dimensions, whereas there are no significant differences between RouteFlow and the vector-field-based method.
Further analysis of mean values and confidence intervals reveals a subtle difference: the vector-field-based method shows slightly lower mental and physical demands and effort than RouteFlow. 
This could be due to the relatively simpler animation paths for individual objects in this method~\cite{wang2017vector}, which might require less cognitive load for tracking compared to the bundled paths in RouteFlow.

\trend: 
The Wilcoxon signed-rank tests indicate that RouteFlow outperforms the focus+context grouping method in terms of mental demand, temporal demand, frustration, and performance dimensions.
This is because the focus+context grouping method demands extra memory and analytical burden to synthesize the movements of multiple groups into the global trend, which often leads to frustration.

\hotspot: 
The Wilcoxon signed-rank tests indicate that RouteFlow outperforms the other two methods on the five dimensions, except for effort, where RouteFlow only outperforms the focus+context grouping method.
These advantages align with participants' accuracy in completing the task.
Regarding the effort dimension, participants found the background grids during the animation helpful, as they reduced the effort needed to observe and locate movement.

\noindent\textbf{Preference}.
We also collected participants' preferences for the three methods in three tasks, as summarized in Fig.~\ref{fig:rank}.

\track: Participants' opinions varied. 
Eight participants preferred RouteFlow, and seven preferred the vector-field-based method.
This was slightly different from the overall subjective workload ratings, where participants thought RouteFlow brought a bit more mental and physical burden and effort.
We noticed that P10 and P15 performed equally accurately with RouteFlow and the vector-field-based method but reported higher subjective performance and preference for RouteFlow.
They commented that objects in RouteFlow were \myquote{less occluded}, making them \myquote{more convinced.}

\trend:
13 participants preferred RouteFlow.
Specifically, P1 appreciated that RouteFlow gradually revealed the global trend along with the group movement, just like \myquote{painting with a brush.}
In contrast, P6 preferred the focus+context grouping method, explaining \myquote{It is easier to connect the movements of different groups into a global trend.}
P10 liked the vector-field-based method and noted that \myquote{I need to memorize the positions objects pass through to identify the global trend. [The vector-field-based method] requires less memory burden, as I can observe objects moving simultaneously in the global trend.}

\hotspot: 
All participants ranked RouteFlow at the top, indicating a strong advantage of our method. 
Their reasons mainly focused on the clear and noticeable animated transitions, showing groups that were converging and diverging.

\setcounter{figure}{12}
\begin{figure}[h]
  \centering
  \includegraphics[width=1.0\linewidth]{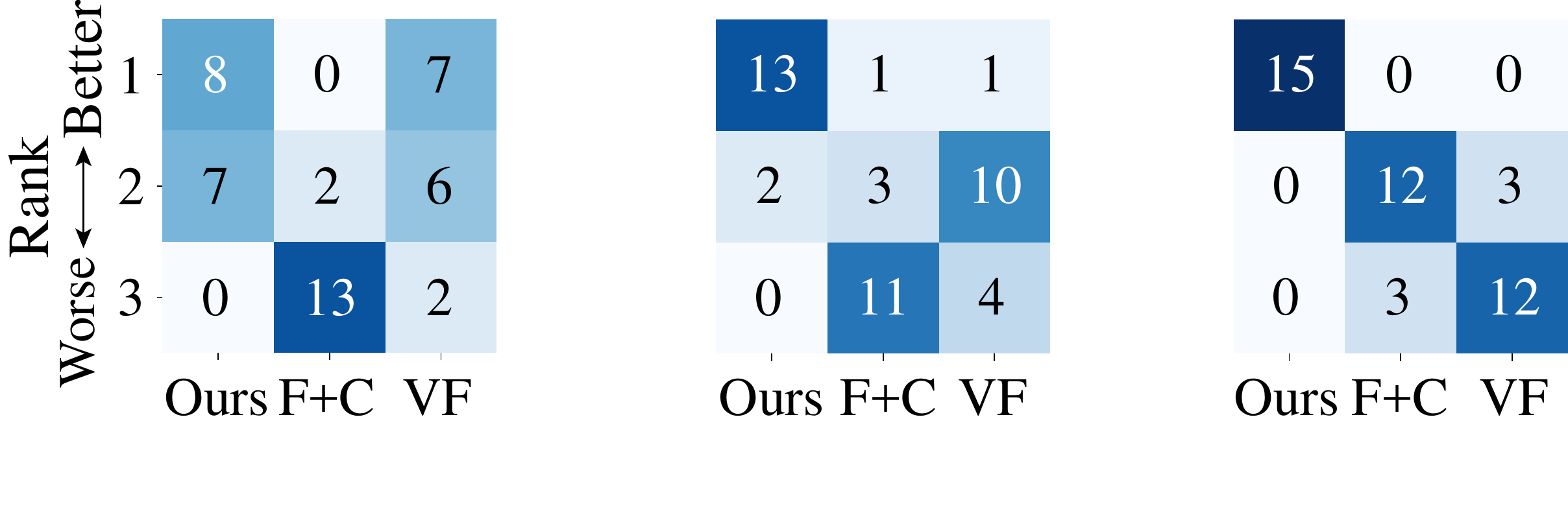}
  \put(-212,3){(a) Task T1}
  \put(-126,3){(b) Task T2}
  \put(-46,3){(c) Task T3}
  \caption{Participants' preference ranks of different methods on three tasks.
  Smaller ranks indicate stronger preferences.}
  \Description{Participants' preference rank of different methods in three tasks.}
  \label{fig:rank}
\end{figure}



\section{Discussion and Future Work}
\label{sec:discuss}
As demonstrated by the results from both the quantitative evaluation and user study, the primary benefit of RouteFlow lies in its capacity to effectively reveal the global trend and identify local hotspots.
It also maintains comparable performance in tracking objects.
The participants generally gave positive feedback on its usability.
Nonetheless, they also pointed out several limitations that deserve further investigation.

\noindent\textbf{Introducing additional visual channels.}
RouteFlow delivers an elaborate animation by planning animation paths and generating a compact and non-occluded object layout. 
In the user study, P10, who majored in information design, suggested including more visual channels to enhance the expressiveness of animations.
For instance, using different colors and brightness can help users track the targeted objects~\cite{hu2016spot}.
Moreover, objects can be encoded in varying sizes based on their importance, making them more easily distinguishable.
A recent study has also validated that the variations in object brightness and size have a positive impact on perceiving groups~\cite{chalbi2020common}.
In the future, we can apply a dynamic color palette to our generated object layout to improve the animation~\cite{chen2024dynamic}.

\noindent\textbf{Interactive animation adjustment.}
Our method automatically generates animations based on the input trajectories, easing the effort required for manual design. 
On top of the automated method, RouteFlow can be enhanced by supporting users to interactively design and refine animations, addressing their unique needs~\cite{wang2017vector}.
For example, users can directly drag to refine the animation paths or modify the corresponding object groups and layout, and our method can propagate the changes to the whole animation. 
Therefore, users do not need to manipulate each object in detail.
Besides, our method can be integrated with dynamic parameter adjustment to balance different types of forces in our hierarchical edge bundling for animation path generation. 
As such, we can analyze data features such as trajectory density, local hotspot distribution, or occlusion levels and adjust parameters on targeted datasets or areas to identify different local hotspots. 
Users can also interactively specify parameters on demands.

\noindent\textbf{Supporting more trajectory patterns.}
While our current method effectively highlights the global trend and local hotspots, trajectory data often contains other patterns that can provide valuable insights~\cite{zheng2015trajectory}.
For instance, periodic patterns represent movements that repeat at regular intervals, such as daily commutes or seasonal migrations~\cite{cao2007periodic}.
Anomalies refer to movements that deviate significantly from the norm, representing unusual or rare behaviors, such as a migratory bird taking an abnormal path due to environmental disruptions~\cite{liu2014fraud,liu2011anomaly}.
A practical solution is integrating existing pattern detection techniques~\cite{chandola2009anomaly} and designing animations to communicate these patterns effectively to users.
For example, trajectories that exhibit periodic patterns could be animated with pulsating effects or cyclic color changes~\cite{aigner2011visualization}, clearly highlighting their repetitive nature over time.

\noindent\textbf{Scalability.}
RouteFlow faces scalability issues due to both algorithmic capabilities and visual constraints.
From the algorithmic perspective, RouteFlow can process hundreds of moving objects in real time.
When scaled to one thousand objects, RouteFlow completes processing in around 10 seconds, making real-time animation impractical at this scale.
From the visual perspective, the number of displayed objects is constrained by both screen space and human perception.
Especially, previous studies have shown that people can only track a limited subset of moving objects~\cite{feria2012effects, franconeri2010tracking},
making human perception a more limiting factor.
A potential solution to address these limitations is to combine a sampling method with our animation method. 
The key challenge is to identify a set of representative samples that effectively capture both the global trend and the local hotspots.



\section{Conclusion}
In this paper, we present RouteFlow, a trajectory-aware animated transition method to enhance the analysis of the movement trend and object tracking in the animation process. 
By analogizing animation paths to bus routes and the object layout to the seat allocation, RouteFlow offers clear and smooth animations of moving objects along their trajectories.
The key feature of RouteFlow lies in its balanced depiction of the global trend and local hotspots, coupled with its ability to minimize occlusion.
This balance improves users' capability to track movements and understand complex interactions within the objects.
The quantitative evaluation and user study further validate that RouteFlow performs better than existing methods in identifying the global trend and locating local hotspots.


\begin{acks}
Duan Li, Xinyuan Guo, and Shixia Liu are supported by the National Natural Science Foundation of China under grants U21A20469, 61936002 and in part by Tsinghua University-China Telecom Wanwei Joint Research Center.
Lingyun Yu is supported by the National Natural Science Foundation of China under grant 62272396.
The authors would like to thank Jiangning Zhu, Zhen Li, Jiashu Chen, Yukai Guo, and Prof. Weikai Yang for their valuable contributions to the discussions and comments.\looseness=-1
\end{acks}

\bibliographystyle{ACM-Reference-Format}
\bibliography{reference}

\clearpage

\appendix


\section{Quantitative Experiment on Edge Bundling}
\label{sec:appendixA}


\begin{table}[b]
  \centering
  \caption{Comparison between different edge bundling algorithms, including divided edge bundling (DEB)~\cite{selassie2011divided}, multilevel agglomerative edge bundling (MAEB)~\cite{gansner2011MINGLE}, and our algorithm (RouteFlow). Lower values are better for both metrics.\looseness=-1
  }
  \label{tab:bundling_result} 
  \resizebox{0.5\textwidth}{!}{
    \begin{tabular}{c|p{7mm}cccccccccccccc}
       \toprule
        \multirow{2}{*}{\textbf{Dataset}}
        & \multicolumn{3}{c}{Deviation}
        & \multicolumn{3}{c}{Ink ratio}\\
        \cmidrule(lr){2-4} \cmidrule(lr){5-7}
        & {DEB} & {MAEB} & {RouteFlow}
        & {DEB} & {MAEB} & {RouteFlow} \\
        \midrule

        Taxi~\cite{yuan2011drive, yuan2010tdrive} &
        0.023 &	0.017 &	\textbf{0.012} &
        0.387 & \textbf{0.349} & 0.378 \\

        BirdMap~\cite{BirdMap} & 
        0.048 &	0.035 &	\textbf{0.016} &
        0.237 &	\textbf{0.235} &	0.239 \\

        Railway~\cite{railway} &
        0.038 &	0.031 &	\textbf{0.018} &
        0.243 &	0.309 &	\textbf{0.234} \\

        MEIBook~\cite{mei} & 
        0.034 &	0.018 & \textbf{0.017} &
        0.397 &	\textbf{0.327} & 0.388 \\

        OpenSkyAirline~\cite{opensky} &
        0.043 & \textbf{0.014} & \textbf{0.014} &
        0.393 & 0.342 & \textbf{0.339} \\

        US Migration~\cite{holten2009force} & 
        0.045 &	0.023 &	\textbf{0.019} &
        0.432 &	0.272 &	\textbf{0.250} \\

        DanishAIS~\cite{DanishAIS} &
        0.071 &	0.076 &	\textbf{0.024} &
        0.213 &	\textbf{0.201} & 0.252 \\

        \midrule
        
        \textbf{Average} & 
        0.043 &	0.030 &	\textbf{0.017} &
        0.329 &	\textbf{0.291} &	0.296 \\ 
        
        \bottomrule
    \end{tabular}
    }
\end{table}


We conducted a quantitative experiment to compare the effectiveness of our trajectory-driven path generation algorithm with two representative edge bundling algorithms.
The experiment aims to assess 1) the deviation of the bundled paths from the original ones, and 2) the efficiency in reducing the total path length. 
The same datasets used in Sec.~\ref{sec:quantitative} were employed for this experiment.

\noindent\textbf{Baseline algorithms.} We selected two representative edge bundling algorithms for comparison.
The first, divided edge bundling (DEB)~\cite{selassie2011divided}, employs the attraction and spring forces to bundle edges with similar positions and directions.
The second, multilevel agglomerative edge bundling (MAEB)~\cite{gansner2011MINGLE}, hierarchically bundles similar edges to minimize the total path length.
For both algorithms, we used the default parameters reported in their papers.

\noindent\textbf{Evaluation criteria.}
We adopted two metrics, deviation and ink ratio~\cite{wallinger2022edgepath}, which assess the deviation from the original paths and the efficiency in reducing the total path length, respectively. 
Let the original path set be $S=\{s_1,\dots, s_n\}$ and the bundled path set be $S'=\{s_1',\dots, s_n'\}$.

\underline{Deviation} measures the misalignment of the bundled paths with the original ones.
Following the common practice~\cite{Francois2011globaldtw,Tao2021comparative}, the misalignment of a bundled path $s_i'$ and its original path $s_i$ is measured by their dynamic time warping distance DTW($s_i$, $s_i'$).
The deviation is then calculated as the average dynamic time warping distance across all pairs of bundled and original paths:
\begin{equation*}
\text{Deviation}(S,S') = \frac{1}{n}\sum^n_{i=1}\text{DTW}(s_i,s_i')
\end{equation*}

\myparagraph{Ink ratio} measures the efficiency in reducing the total path length.
It is calculated as the ink of the bundled path set $S'$ divided by that of the original path set $S$:
\begin{equation*}
\text{Ink ratio}(S,S') = \frac{\text{Ink}(S')}{\text{Ink}(S)},
\end{equation*}
where $\text{Ink}(S)$ is the number of pixels occupied by the paths in $S$~\cite{wallinger2022edgepath}.\looseness=-1

\setcounter{figure}{13}
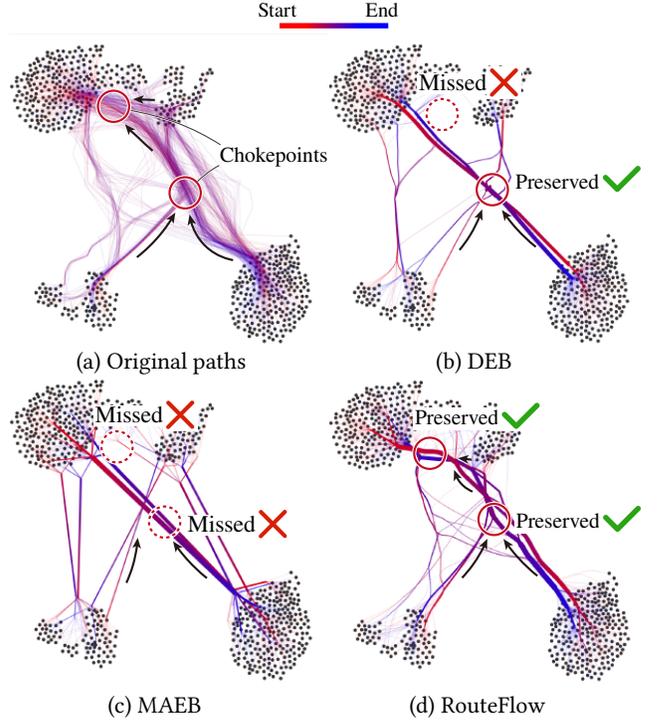
\begin{figure}[t]
  \centering
  \includegraphics[width=1.0\linewidth]{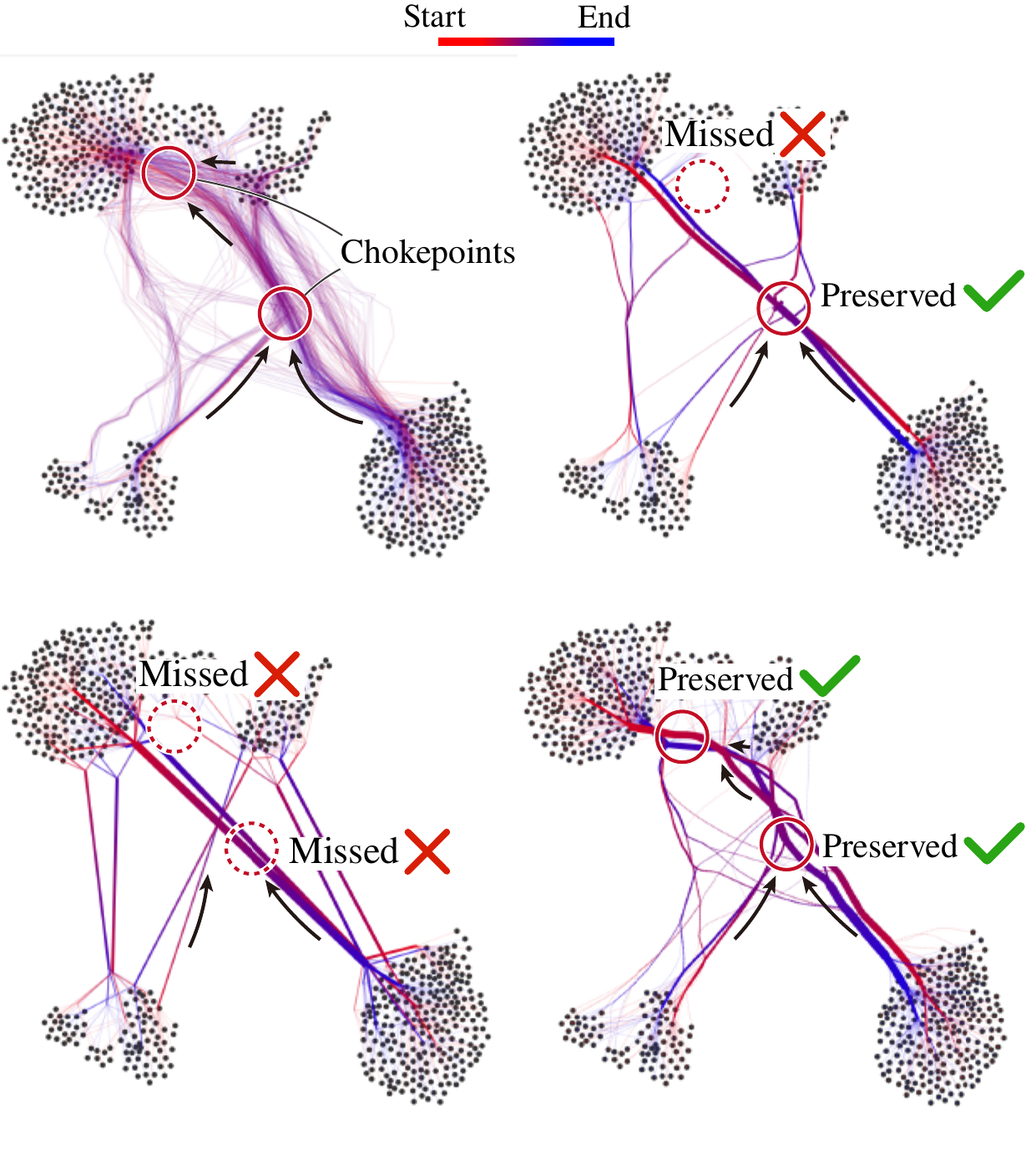}
  \put(-216,133){(a) Original paths}
  \put(-80,133){(b) DEB}
  \put(-204,3){(c) MAEB}
  \put(-90,3){(d) RouteFlow}
  \caption{
  Original paths and the edge bundling results on the DanishAIS dataset.
  }
  \label{fig:danish}
  \Description{Original paths and the edge bundling results on the DanishAIS dataset.}
\end{figure}

\noindent\textbf{Results}

Table~\ref{tab:bundling_result} shows the comparison results on seven datasets. 
These results indicate that our algorithm achieves the lowest deviation and performs comparably with the baseline algorithms in terms of ink ratio.

\setcounter{figure}{14}
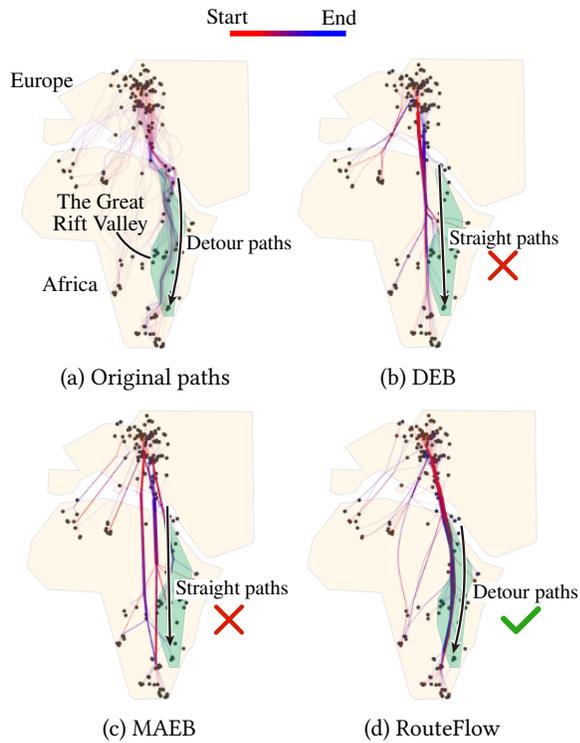
\begin{figure}[t]
  \centering
  \setlength{\abovecaptionskip}{1.2mm}
  \includegraphics[width=0.9\linewidth]{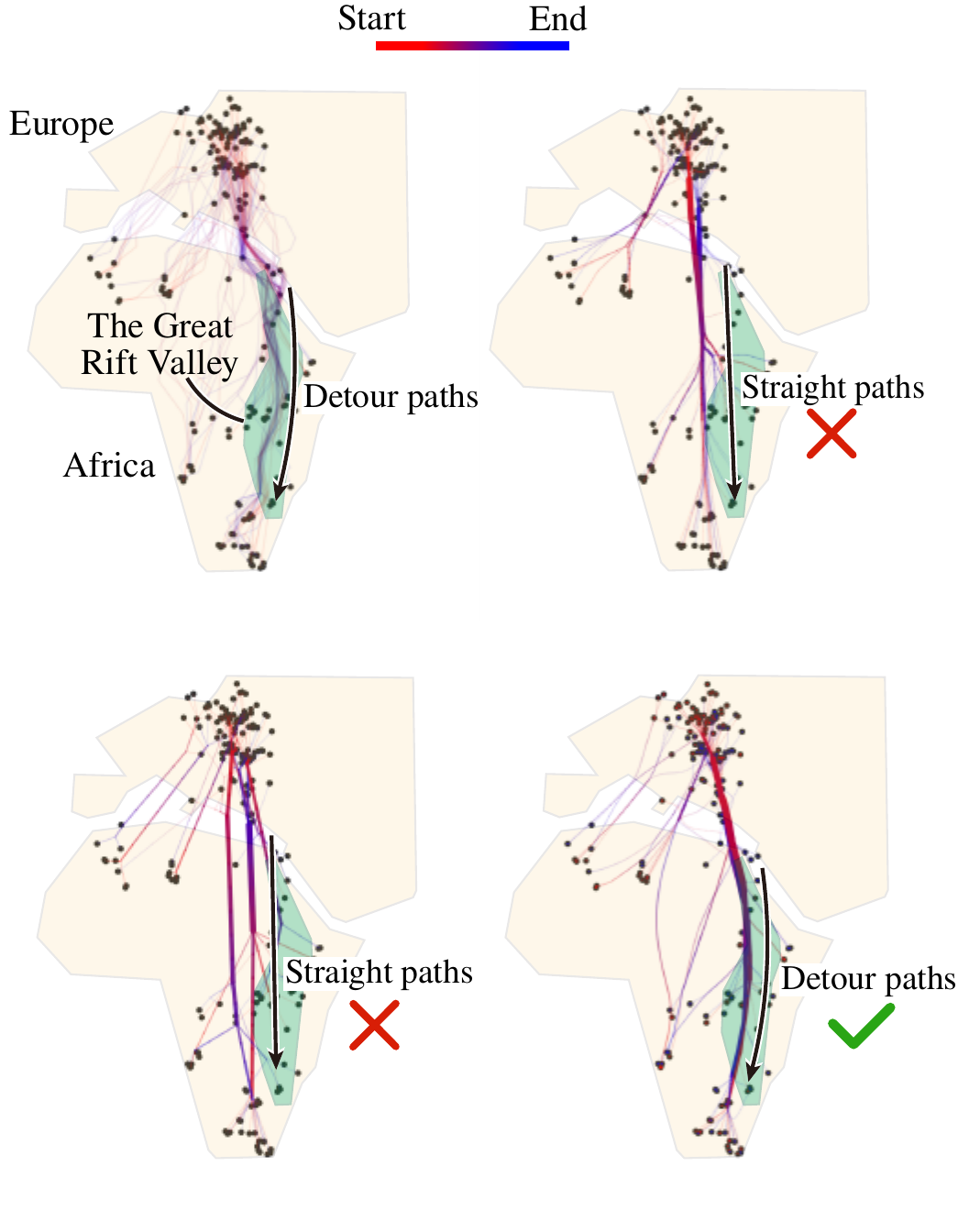}
  \put(-196,136){(a) Original paths}
  \put(-75,136){(b) DEB}
  \put(-180,3){(c) MAEB}
  \put(-81,3){(d) RouteFlow}
  \caption{
  Original paths and the edge bundling results on the BirdMap dataset.
  }
  \label{fig:birdmap}
  \Description{Original paths and the edge bundling results on the BirdMap dataset.}
\end{figure}
\underline{Deviation}.
The baseline algorithms perform worse in terms of deviation because they do not explicitly preserve the original paths and often excessively aggregate paths. 
This leads to larger deviation from the original paths and hinders the identification of local hotspots.
For example, as shown in Fig.~\ref{fig:danish}(a), the DanishAIS maritime transportation dataset contains two chokepoints that significantly impact maritime traffic efficiency, making them strategic hotspots for route optimization.
However, the baseline algorithms fail to preserve these chokepoints (Figs.~\ref{fig:danish}(b) and (c)).
In contrast, RouteFlow reduces the deviation from the original paths using an anchor force and effectively preserves these two chokepoints (Fig.~\ref{fig:danish}(d)).
This preservation enables more accurate and effective route optimization.

\underline{Ink ratio}.
Out of the seven datasets, MAEB achieves the lowest ink ratio in four, while RouteFlow achieves the lowest in three.
MAEB's better performance is due to its focus on optimizing ink ratio, but this comes at the cost of higher deviation from the original paths.
In contrast, RouteFlow balances ink ratio and deviation, sacrificing a small amount of ink ratio to better align the bundled paths with the original ones.
This balance is important in generating clear and reliable animation.
For example, as shown in Fig.~\ref{fig:birdmap}(a), migratory birds in the BirdMap dataset take detour paths along the Great Rift Valley to access water or suitable habitats during their journey from Europe to Africa.
These detour paths are critical to understanding the birds' movement patterns in the context of geographical features.
However, MAEB distorts the original paths and bundles them into straight paths to minimize the ink ratio, which obscures the birds' actual movement patterns (Fig.~\ref{fig:birdmap}(c)).
Although DEB does not explicitly minimize the ink ratio, it also bundles the paths into straight paths and obscures the movement patterns in this example (Fig.~\ref{fig:birdmap}(b)).
In contrast, RouteFlow effectively bundles the paths while preserving the detour paths (Fig.~\ref{fig:birdmap}(d)).
This provides a clearer and more reliable visual summary of the birds' movements.\looseness=-1

\section{Detailed Animation Result Analysis}
\label{sec:appendixB}
To facilitate interpreting the magnitudes of the measured values, Fig.~\ref{fig:metrics} compares the animation results with different measured values, which are generated by three methods:  RouteFlow, the focus+context grouping method (F+C) and the vector-field-based method (VF).
These methods are the same as in Sec.~\ref{sec:quantitative}. 
Fig.~\ref{fig:metrics}(a) shows the positions of all objects, the same as in Fig.~\ref{fig:quan-result}. 
For easy comparison, we select a subset of 10 objects in a group (with the orange contour), and calculate their values for each metric in this frame (Figs.~\ref{fig:metrics}(c)-(e)), except for overall occlusion, which is calculated on all objects (Fig.~\ref{fig:metrics}(b)).

\setcounter{figure}{15}
\begin{figure}[t]
  \centering
  \includegraphics[width=\linewidth]{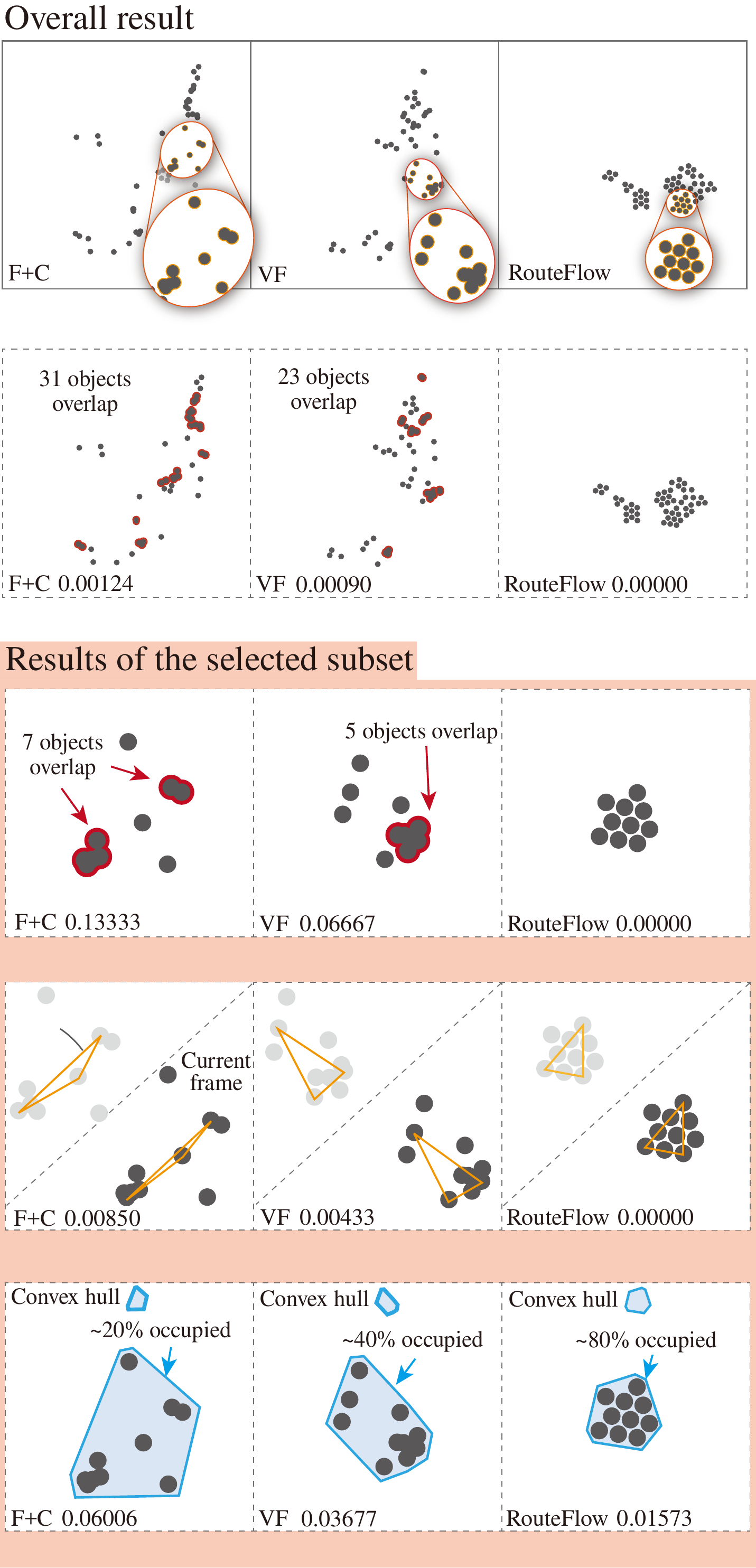}
  \put(-207,394){(a) A subset of 10 objects for detailed analysis}
  \put(-160,300){(b) Overall occlusion}
  \put(-172,191){(c) Within-group occlusion}
  \put(-151,97){(d) Deformation}
  \put(-147,3){(e) Dispersion}
  \caption{Comparisons of the three animation results with different metric values, which are generated by the three methods.
  Here, (b)-(e) show the different metric values of the 10 objects.
  }
  \label{fig:metrics}
  \Description{Illustration of how the metric values change with different object positions.}
\end{figure}

\myparagraph{Occlusion}.
Figs.~\ref{fig:metrics}(b) and (c) compare the animation results generated by the three methods in terms of the overall and within-group occlusion.
RouteFlow prevents overlap between objects when they move together as a group, thereby achieving an occlusion score of $0.00000$ for both metrics at this frame.
However, overlaps remain unavoidable at local hotspots where objects converge and diverge.
In contrast, the focus+context grouping method and the vector-field-based method result in object overlap to different degrees.
For example, the focus+context grouping method achieves an overall occlusion score of $0.00124$, with 31 out of 51 objects overlapping, and a within-group occlusion score of $0.13333$, with $7$ of the $10$ objects overlapping.

\underline{Deformation}.
Fig.~\ref{fig:metrics}(d) compares the animation results generated by the three methods in terms of deformation. 
This metric measures the changes in distance between objects from the previous frame (grey) to the current frame (dark grey).
RouteFlow organizes the objects that move together into a cohesive group, with their relative positions changing only at local hotspots.
As a result, the deformation score is $0.00000$, indicating no changes in the distance between objects.
In contrast, the focus+context grouping method and the vector-field-based method do not explicitly manage the object layout, leading to larger deformation scores.
For example, the focus+context grouping method achieves a deformation score of $0.00850$, indicating that the average distance between objects has changed by approximately $1.46$ times the object's radius.

\underline{Dispersion}.
Fig.~\ref{fig:metrics}(e) compares the animation results generated by the three methods in terms of dispersion.
RouteFlow closely groups the objects that move together, resulting in the lowest dispersion score of $0.01573$, with the objects occupying approximately $80\%$ of the space within their convex hull.
In contrast, the focus+context grouping method and the vector-field-based method exhibit higher dispersion scores.
For example, the focus+context grouping method achieves a dispersion score of $0.06006$, with the objects occupying approximately $20\%$ of the space within their convex hull.

\section{User Study Settings}
\label{sec:appendixC}
\subsection{Synthetic Dataset Generation}
As illustrated in Fig.~\ref{fig:datageneration}, our dataset generation process involves three steps: generate the global trend, determine local hotspots, and create trajectories.

Since the global trend can be depicted by representative trajectories~\cite{zheng2015trajectory}, we generate a smooth trajectory using B-splines as the ground truth.
We include one or two bends in these B-splines, leading to 2 types of the global trend in our dataset. 
To achieve this, we randomly select two points as the start and end points and sample one (for one bend) or two (for two bends) intermediate points as B-spline control points within the region between them.
To avoid highly curved trajectories, we ensure that the angle between any three consecutive points (start, control points, and end) is greater than 135 degrees.

Next, to determine local hotspots, we randomly sample points along the generated B-spline and designate these points as ground truth for local hotspots.
Feedback from our pilot study indicates that 
to balance complexity and diversity, the number of local hotspots should be limited to two or three, with each dataset containing at least one of the local hotspot types: convergence or divergence.
Each local hotspot is then randomly assigned as either a converging or diverging point, resulting in three possible assignments: 1 convergence + 1 divergence, 2 convergences + 1 divergence, and 1 convergence + 2 divergences.
Considering the types of the global trend and the local hotspot assignments, we generate 6 dataset types (2 types of the global trend $\times$ 3 local hotspot assignments).
To simplify the evaluation, we 1) limit the number of branches at the local hotspots to 2; 2) select one B-spline as the global trend and generate another branch at the local hotspots along this B-spline; and 3) avoid crossing between the two branches of each local hotspot.

Finally, we generate the trajectories for the objects.
Following the user feedback and the setting of previous studies~\cite{zheng2018focus+}, we set the number of trajectories to 30.
The trajectories are generated by adding random perturbations to the global trend and the branches of the local hotspots.

\subsection{Method Counterbalance}
In our user study, we counterbalanced the order of different methods.
We divided 15 participants into five groups, with three participants in each group. 
Within each group, we used an expanded Latin square and applied a cyclic shift to the method order for each participant.
The three methods are denoted as A, B, and C.
In the test scenario, the order in which these methods were presented to the participants in each group was as follows:

\begin{itemize}
    \item A B C, B C A, C A B, A B C, B C A, C A B, A B C, B C A, C A B, A B C, B C A, C A B
    \item B C A, C A B, A B C, B C A, C A B, A B C, B C A, C A B, A B C, B C A, C A B, A B C
    \item C A B, A B C, B C A, C A B, A B C, B C A, C A B, A B C, B C A, C A B, A B C, B C A
\end{itemize}

\end{document}